\documentclass{comnet}

\usepackage{amsmath}
\usepackage{amsthm}
\usepackage{cite}
\usepackage{adjustbox}
\usepackage{caption}
\usepackage[multiple]{footmisc}
\usepackage{booktabs}
\usepackage{float}
\usepackage{afterpage}
\usepackage{epstopdf}
\usepackage{tablefootnote}
\usepackage[numbers]{natbib}

\begin{document}

\title{Topological-temporal properties of evolving networks}

\shorttitle{Topological-temporal properties of Evolving Networks} 
\shortauthorlist{A.Ceria \textit{et al.}} 

\author{
\name{Alberto Ceria$^*$}
\address{Faculty of Electrical Engineering,
Mathematics, and Computer
Science, Delft University of
Technology, Mekelweg 4, 2628 CD,
Delft, The Netherlands}
\name{Shlomo Havlin}
\address{Department of Physics, Bar Ilan University, Ramat Gan, Israel; Department of Physics, Boston
University, Boston, MA, USA and Tokyo Institute of Technology, Yokohama, Japan}
\name{Alan Hanjalic}
\address{Faculty of Electrical Engineering,
Mathematics, and Computer
Science, Delft University of
Technology, Mekelweg 4, 2628 CD,
Delft, The Netherlands}
\and
\name{Huijuan Wang}
\address{Faculty of Electrical Engineering,
Mathematics, and Computer
Science, Delft University of
Technology, Mekelweg 4, 2628 CD,
Delft, The Netherlands\email{$^*$Corresponding author: A.Ceria@tudelft.nl}}}
\maketitle

\begin{abstract}
{Many real-world complex systems including human interactions can be represented by temporal (or evolving) networks, where links activate or deactivate over time. Characterizing temporal networks is crucial to compare different real-world networks and to detect their common patterns or differences. A systematic method that can characterize simultaneously the temporal and topological relations of the time specific interactions (also called contacts or events) of a temporal network, is still missing. In this paper, we propose a method to characterize to what extent contacts that happen close in time occur also close in topology. Specifically, we study
the interrelation between temporal and topological properties of the contacts
from three perspectives: (1) the correlation (among the elements) of the activity time series which records the total number of contacts in a network that happen at each
time step; (2) the interplay between the topological distance and time difference of two arbitrary contacts; (3) the temporal correlation of contacts within the local
neighborhood centered at each link (so called ego-network) to explore whether such contacts that happen close in topology are also close in time. By applying our method to 13 real-world
temporal networks, we found that temporal-topological correlation of contacts
is more evident in virtual contact networks than in physical contact networks. This could be due to the lower cost and easier access of online communications than
physical interactions, allowing and possibly facilitating social contagion, i.e., interactions of one individual may influence the activity of its neighbors. We also identify different patterns between virtual and physical networks and among physical contact networks at, e.g., school and workplace, in the formation of correlation in local neighborhoods. 
Patterns and differences detected via our method may further inspire the development of more realistic temporal network models, that could reproduce jointly temporal and topological properties of contacts.}
{Evolving networks, Structural analysis of networks, Social, socio-economic and political networks}

\end{abstract}

\section{Introduction}

Complex systems can be represented as networks, where nodes and links represent the components of a system and their interactions respectively. In a temporal or evolving network \citep{holme2012temporal,holme2015modern}, the network topology changes over time, or equivalently, pairs of nodes interact at specific time stamps. Such time-stamped interactions between nodes are called contacts or events.
Early work on evolving networks and their characterization methods have mostly focused on either temporal \citep{goh2008burstiness,eckmann2004entropy,oliveira2005darwin,candia2008uncovering,johansen2004probing} or topological \citep{barabasi2003scale,barabasi2009scale,newman2003structure,boccaletti2006complex,barrat2004architecture,onnela2007structure} dimension separately but rarely on combining both \citep{brot2016evolution,kikas2013bursty,paranjape2017motifs,kovanen2011temporal,karsai2012correlated,pan2011path}. Regarding the topological aspect, the aggregated networks, where two nodes are connected if they have at least one contact or interaction, have been characterized using classical static network analysis methods. Scaling properties such as a scale-free degree distribution have been observed in many real networks \citep{barabasi2003scale,barabasi2009scale,newman2003structure,boccaletti2006complex}. From the perspective of time dimension, it has been found that individuals tend to execute actions like contacts in bursts within a short time duration and such high activity periods are separated by relatively long inactive ones. The approximate scale-free distribution of the inter-event times of contacts of a node or of a system, the so-called burstiness, seems to be common in real-world temporal networks \citep{goh2008burstiness,barabasi2005origin,eckmann2004entropy,oliveira2005darwin,vazquez2007impact,candia2008uncovering,johansen2004probing}. The temporal correlation of the events of a network has been measured by e.g. auto-correlation \citep{rybski2009scaling} and the distribution of the number of contacts in a bursty period, the so-called event train. \citep{karsai2012universal}.

Recent studies have started to characterize both the topological and temporal properties together. It has been observed that events of addition and removal of links by users do not occur sporadically at random nodes but rather occur in brief bursts in time and locally in topology, on both an online blogging platform and Skype \citep{brot2016evolution,kikas2013bursty}. Temporal motifs are sets of contacts among a small number of nodes conforming to a specific pattern in topology and time ordering as well as a specific duration of time. The occurrence of diverse temporal motifs has been used to characterize and to classify evolving networks \citep{paranjape2017motifs,kovanen2011temporal}. Karsai et al.  \citep{karsai2012correlated} characterized the sequence of contacts between each node and its neighbours using the distribution of the number of contacts in a bursty period, which is also called the event train size. However, it has been shown that bursty trains are usually formed by contacts between pair of nodes instead of in the aforementioned neighborhood of a node.

However, systematic methods to characterize simultaneously the temporal and topological properties of contacts/events to better understand real-world networks' differences and similarities are still missing. In this work, we aim to develop methods to characterize to what extent contacts that happen close in time (topology) are also close in topology (time). Specifically, we characterize the relationship between temporal and topological properties of the contacts in real evolving networks from the following three perspectives: (a) The auto-correlation of the activity time series which records the total number of contacts in a network that happen at each time step; (b) The interplay between the topological distance and temporal delay of two contacts; (c) The temporal correlation of contacts within local neighborhoods beyond a node pair. These perspectives characterize simultaneously both the temporal and topological interrelations of contacts from a global level to a more granular level. In order to be able to characterize and compare real-world networks, normalization and three control network randomizations have been designed in our characterization methods. We apply our method to $13$ real-world physical and virtual contact networks. We find that the temporal and topological correlation tends to be more evident in virtual contact networks compared to physical contact networks. This is likely because the online communications, which are of lower cost and easier to perform than physical contacts, allows and possibly facilitates social contagion, i.e. the interaction of one individual to influence the activity of its neighbors. 
At the local neighborhood centered at each link, we observe long trains of events, i.e., consecutive activations of links in the neighborhood. In physical contact networks, the number of distinct links whose activations contribute to a train seems to reflect the spatial constrains of interactions. For example, 
the number of distinct links activated in a train is larger (smaller) in a primary school (workplace) where contacts are less (more) constrained in space.


The detected patterns and differences could further guide the development of evolving network models, pushing the boundary of temporal network models towards reproducing jointly realistic temporal and topological properties. Moreover, temporal network properties influence the dynamic process which unfolds on the network \citep{zhan2019information,zhan2019suppressing,pfitzner2013betweenness,miritello2011dynamical,kivela2012multiscale,scholtes2014causality,williams2019auto,backlund2014effects,pan2011path,parshani2010dynamic,horvath2014spreading,delvenne2015diffusion}. The temporal and topological correlation in an evolving network discovered using our methods could possibly better explain the dynamic process than topological property or temporal property alone.

\section{Definitions}
\subsection{Representation of a temporal network}
A network whose topology vary over time is called a temporal or evolving network.
It can be represented by $\mathcal{G = (N,L)}$, where $\mathcal{N}$ is the set of nodes (with size $|\mathcal{N}| = N$), $\mathcal{L} = \{\mathcal{\ell} (i,j,t), t \in [0,T) , i,j \in \mathcal{N}\}$ is the set of contacts, and each element $\mathcal{\ell} (i,j,t)$ indicates that a contact or an interaction between node $i$ and $j$ occurs at  time $t$. 
A temporal network can also be represented by a 3 dimensional adjacency matrix $\mathcal{A}_{N \times N \times T}$  whose elements $\mathcal{A}(i,j,t)=1$ or $\mathcal{A}(i,j,t)=0$ represent, respectively, the presence or the absence of a contact between node $i$ and $j$ at time $t$.

We consider undirected temporal networks, where $\mathcal{\ell} (i,j,t) = \mathcal{\ell} (j,i,t)$ and $\mathcal {A}(i,j,t) = \mathcal {A}(j,i,t)$.
By aggregating the contacts between each node pair over the whole observation time $[0,T-1]$ one obtains the time aggregated network $G_W = (\mathcal{N},\mathcal{L}_W)$. The aggregated network is static: two nodes $i$ and $j$ are connected, i.e., $e(i,j) \in \mathcal{L}_W$, if there is at least one contact between $i$ and $j$ over the observation time $[0,T-1]$. The adjacency matrix of the unweighted aggregated network is denoted by $A_{N \times N}$ whose element $A(i,j)=1$ or $A(i,j)=0$ depending whether $i$ and $j$ are connected or not. Each link $e(i,j)$ in $\mathcal{L}_W$ can be further associated with a weight $W(i,j)$, which represents the total number of contacts between $i$ and $j$ over the time window $[0,T-1]$. The corresponding weighted adjacency matrix $W_{N \times N}$ has elements $W(i,j) = \sum_{t=0}^{t=T-1}\mathcal {A}(i,j,t)$.

\subsection{Temporal distance and topological distance between two contacts}
\label{sec:temp_spac_distance}
The contacts between two arbitrary nodes $i$ and $j$ can be regarded as the activation of the link $e(i,j) \in \mathcal{L}_W$ at the corresponding time stamps. The activity between $i$ and $j$ can be represented by a time series $X_{ij} =\{x_{ij}(t)=\mathcal {A}(i,j,t), t \in [0,T-1]\}$. The link $e(i,j)$ is active at time $t$ if there is a contact between $i$ and $j$ at time $t$, i.e. $x_{ij}(t) = \mathcal {A}(i,j,t)=1$. The total number of contacts in a network at each time stamp $Y=\{y(t) = \sum_{i,j \in \mathcal{N}, i<j}x_{ij}(t), t\in [0,T-1]\}$ reflects the global activity of the temporal network over time. 
The temporal distance between two contacts $\ell(i,j,t)$ and $\ell(k,l,s)$ is $\mathcal{T}(\ell(i,j,t), \ell(k,l,s)) = |t-s|$.

The topological distance, also called hopcount, between two nodes on a static network is the number of links contained in the shortest path between these two nodes. We define the topological distance $\eta(\mathcal{\ell}(i,j,t),\mathcal{\ell}(k,l,s))$ between two contacts $\mathcal{\ell}(i,j,t)$ and $\mathcal{\ell}(k,l,s)$ as the distance $\eta(e(i,j),e(k,l))$ between the corresponding two links $e(i,j)$ and $e(k,l)$ on the unweighted aggregated network, $G_{W}$. It can be derived as follows. The distance between the same link is zero, e.g. $\eta(e(i,j),e(i,j))=0$. The distance between two different links follows
\begin{equation}
\label{eq:equation1}
    \eta(e(i,j),e(k,l))  \newline  = \min_{u\in \{i,j\},\ v\in \{k,l\}\}}(h(u,v)+1)
\end{equation}
where $h(u,v)$ is the distance or hopcount between node $u$ and $v$ on the unweighted aggregated network $G_W$. The distance between two links is thus one plus the minimal distance between two end nodes of the two links. For example $\eta(e(i,j),e(i,k))=1$.
Moreover, the line graph, e.g, $G_W^{L}$ of a network $G_W$ can be constructed by considering each link in $G_W$ as a node, and two nodes are connected in $G_W^{L}$ if the two corresponding links in $G_W$ share a same end node. The distance (\ref{eq:equation1}) between two links in $G_W$ equals the hopcount between their corresponding nodes in the line graph $G_W^{L}$.

\subsection{Network randomization -control methods}
In Section \ref{sec:method}, we will explore diverse temporal-topological properties to understand the temporal and topological interrelations between contacts. However, real-world evolving networks may differ in, e.g., the number of nodes and the number contacts. In order to detect the non-trivial temporal-topological features and their interrelations in real-world networks, we compare each real-world network with its three controlled randomized networks which systematically preserve or remove specific topological and temporal correlation of contacts.

  For a given temporal network $\mathcal{G}$, we introduce three randomized temporal networks $\mathcal{G}^{1}$, $\mathcal{G}^{2}$ and $\mathcal{G}^{3}$ respectively. Consider the set of contacts $\{\ell(i,j,t)\}$ in a temporal network $\mathcal{G}$, where each contact is described by its topological location, i.e., between pair of  nodes $(i,j)$ and its time stamp, $t$. Randomized network $\mathcal{G}^{1}$ is obtained by reshuffling the time stamps among the contacts, without changing the topological locations of the contacts. This randomization does not change the number of contacts between each node pair, only the timing is randomly changed, thus preserving the probability distribution of the topological distance of two randomly selected contacts.  A temporal network can be also considered as an unweighted aggregated network and each link $e(i,j) \in \mathcal{L}_W$ is associated with its activity time series $\{\mathcal {A}(i,j,t), t \in [0,T-1]\}$. Randomized network $\mathcal{G}^{2}$ is obtained by iterating the step where two links are randomly selected from the aggregated network and their time series are swapped. This randomization does not change the distribution of the inter-event time of the activity of a random link, shown in Figures \ref{fig:fig15} (virtual contacts) and \ref{fig:fig16} (physical contacts). The third randomized network $\mathcal{G}^3$ is obtained by swapping the activity time series of two randomly selected links but with the same total number of contacts. This randomization preserves the number of contacts per node pair, the distribution of the inter-event time of contacts between a node pair and the distribution of the topological distance of two randomly selected contacts. The three randomized networks lead to 
the same unweighted aggregated network as the original network $\mathcal{G}$.

\section{Datasets}
\begin{figure}[H]
\includegraphics[width = 0.9\textwidth]{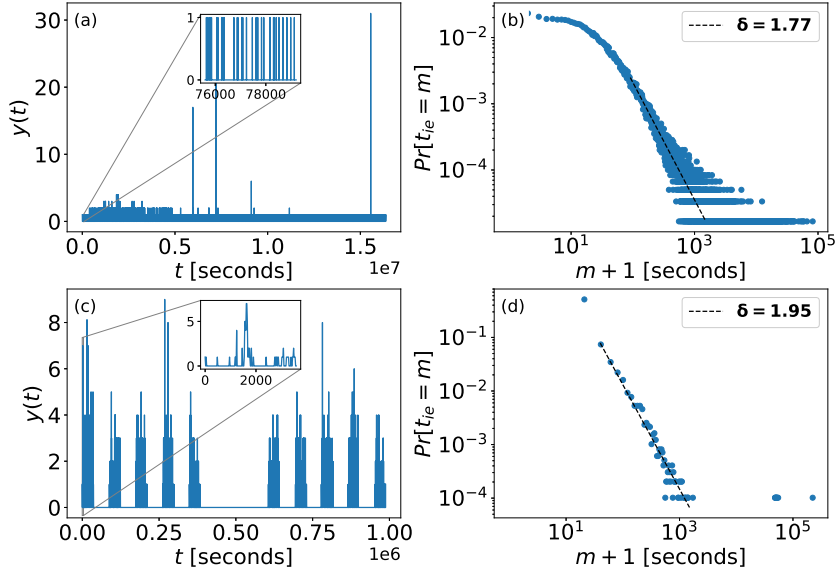}
\caption{Global activity (left) and its inter-event time distribution (right) in (a-b) virtual contact network CM and (c-d) physical contact network WP. The dashed line indicates the slope $\delta$ of the power-law fit and the scaling region, obtained via Clauset's method \citep{clauset2009power}. If the goodness  of the power-law fit is significantly better than the exponential fit, the value of $\delta$ is reported in bold characters \protect\footnotemark.  Time is expressed in seconds. Values of global activity are the total number of contacts occurred in each step of $dt$ seconds. Insets in left figures show global activity for one hour. In WP, long time periods of null global activity correspond to night and weekend periods. These periods correspond to isolated outliers in the global inter-event time distribution with $m>10^{4}s$ and are removed in the data processing.} 
\label{fig:fig1}
\end{figure}

All datasets of temporal networks are obtained from open access websites \footnote{http://www.sociopatterns.org/}\footnote{http://konect.uni-koblenz.de/}\footnote{https://snap.stanford.edu/data/index.html}.
For each dataset, we consider nodes that belong to the largest connected component of the static aggregated network. The corresponding temporal network captures only the contacts between those nodes. Furthermore, we remove the long periods without any contact in the network, corresponding to e.g. night or weekend: we recognized these periods as outliers in the inter-event time \footnote{ The inter-event time $t_{ie}$ is the time interval between the occurrence of two consecutive events. A global activity time series $Y$ with total number of events $k = \sum_{t=0}^T y(t)$  has $k-1$ inter-event times. If two events are contemporary, their corresponding inter-event time is 0.} distribution of the global activity series $Y$ that are far from the bulk.  (see Figure~\ref{fig:fig1}). Finally, multiple contacts between the same pair of nodes at the same time step are accounted as a single contact. Details of the datasets are given in Table~\ref{tab:1}. In the original DNC Mail dataset \footnote{http://konect.uni-koblenz.de/}, more than 96\% of the total contacts forming the largest connected component occur in the last 33 days out of the 982 days. Hence, we include only the contacts of the last 33 days in our DNC Mail data.

\begin{table}[H]
\begin{adjustbox}{width=\columnwidth,center}
\begin{tabular}{@{}llllllllr@{}}
Network     & $N$  &  $|\mathcal{L}_W|$ & $|\mathcal{S}|$ & $|\mathcal{L}|$  & $T$& $dt$     &\textit{contact type}   \\ 
\midrule
DNC Mail Part 2 (DNC\_ 2 *) \citep{kunegis2013konect}    &    1598     &    4085     &    17300     &    30091     &    2861358     &    1     &    \textit{virtual}\\ 
Manufacturing Email (ME*)\citep{emailsRadoslaw}     &    167     &    3250     &    57791     &    82281     &    23430482     &    1     &    \textit{virtual}       \\ 
College Messages (CM*)\citep{panzarasa2009patterns}     &    1892     &    13833     &    58905     &    59789     &    16362751     &    1     &    \textit{virtual}       \\ 

Email EU (EEU*)\citep{leskovec2007graph,paranjape2017motifs}     &    986     &    16025     &    206311     &    324933     &    44719809     &    1     &    \textit{virtual}       \\

Infectious (Infectious)\citep{isella2011s}     &    410     &    2765     &    1392     &    17298     &    1421     &    20     &    \textit{physical}       \\

Primary School (PS)\citep{stehle2011high}     &    242     &    8317     &    3099     &    125771     &    3098     &    20     &    \textit{physical}       \\

High School 2012 (HS2012)\citep{fournet2014contact}     &    180     &    2220     &    11267     &    45047     &    14114     &    20     &    \textit{physical}       \\

High School 2013 (HS2013)\citep{mastrandrea2015contact}     &    327     &    5818     &    7371     &    188504     &    7370     &    20     &    \textit{physical}       \\

Hypertext 2009 (HT2009)\citep{isella2011s}     &    113     &    2196     &    5243     &    20818     &    7226     &    20     &    \textit{physical}       \\

SFHH Conference (SFHH)\citep{cattuto2010dynamics,stehle2011simulation}     &    403     &    9565     &    3508     &    70261     &    3799     &    20     &    \textit{physical}       \\

Workplace 2013 (WP)\citep{genois2015data}     &    92     &    755     &    7095     &    9827     &    17844     &    20     &    \textit{physical}       \\

Workplace 2015 (WP2)\citep{genois2018can}     &    217     &    4274     &    18479     &    78246     &    20946     &    20     &    \textit{physical}       \\

Hospital (Hospital)\citep{vanhems2013estimating}     &    75     &    1139     &    9452     &    32424     &    16026     &    20     &    \textit{physical}       \\

\end{tabular}
\end{adjustbox}
\caption{Basic features of the empirical networks after data processing. The number of nodes ($N = |\mathcal{N}|$), the number of links  in $\mathcal{L}_W$ ($|\mathcal{L}_W|$), the number of snapshots ($|\mathcal{S}|$), the total number of contacts ($|\mathcal{L}|$), the length of the observation time window in time steps ($T$), the time resolution or duration of each time step ($dt$) in seconds and contact type are shown.}
\label{tab:1}
\end{table}

\footnotetext{This evaluation is performed via the likelihood ratio test on power-law and exponential fits. If the test indicate a better performance of the power-law fit with p-value $p<0.05$, then the exponent of power-law fit $\delta$ is reported in bold characters. }
\section{Characterizing topological-temporal properties of evolving networks}
\label{sec:method}
In this Section, we propose a systematic  method to characterize topological-temporal properties of the contacts in an evolving network. In Subsection ~\ref{sec:global_activity} we focus on the characterization of temporal properties, while in Section ~\ref{sec:top_temp_analysis} and \ref{sub:temp_corr_local} we characterize the joint topological and temporal features of contacts.

\subsection{Temporal analysis of global activity}
\label{sec:global_activity}
The time series of global activity $Y=\{y(t),t\in[0,T-1]\}$ records the total number of contacts at each time step $t\in[0,T-1]$. In this Section, we analyze the correlation among the elements of the global activity time series.

\begin{figure}[!h]
    \centering
	\includegraphics[scale = 0.55]{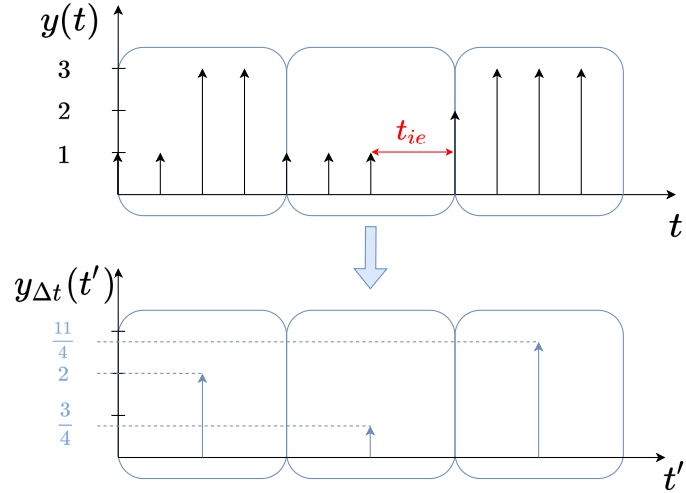}
    \caption{Construction of the aggregated activity series $y_{\Delta t}(t')$ from the global activity time series ${y(t)}$, where $\Delta t = 4$ time steps. In the top sub-figure, we present the event sequence of $y(t)$, where each vertical line indicates the timing of one (or more -depending on thickness) event(s), while $t_{ie}$ is the inter-event time and two events happening at the same time have an inter-event time zero.}
    \label{fig:fig2}
\end{figure}

 We aggregate the global activity at each time bin of duration $\Delta t$ time steps as follows. The time steps $t\in[0,T-1]$ can be divided into a set of non-overlapping consecutive time bins of duration $\Delta t$. The aggregated activity $y_{\Delta t}(t')$ at a time bin $[t'\Delta t, t'\Delta t+\Delta t)$ is the average activity of $y(t)$ within the time bin $[t'\Delta t, t'\Delta t+\Delta t)\}$, as shown in Figure \ref{fig:fig2}. Given bin duration $\Delta t$, the aggregated time series of activity is  $Y_{\Delta t}=\{y_{\Delta t}(t'),  0\leq t'\leq \left\lfloor\dfrac{T-1}{\Delta t}\right\rfloor -1\}$.
\begin{figure}[!h]
    \centering
    \includegraphics[scale=0.75] {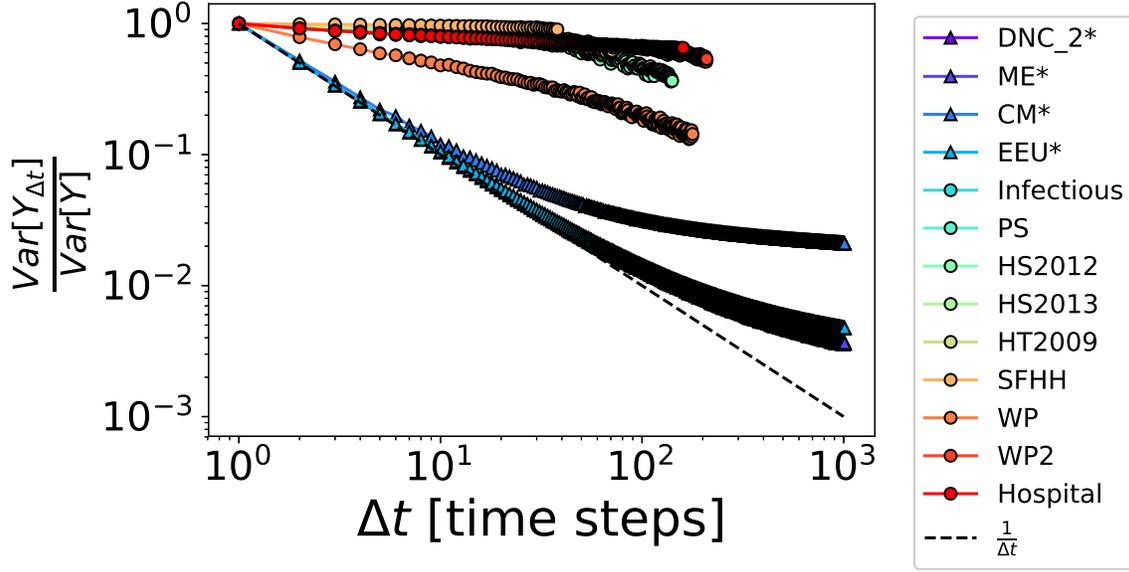}
    \caption{The normalized variance $ \frac{Var[Y_{\Delta t}]}{Var[Y]}$ as a function of the aggregation resolution $\Delta t$.
    Circles correspond to physical contact temporal networks, triangles correspond to virtual contact networks (online messages and mail), while the black dashed line ($\frac{\text{Var}[Y_{\Delta t}]}{\text{Var}[Y]}= \frac{1}{\Delta t}$) represents the uncorrelated curve. The resolution $\Delta t$ is in units of time steps.}
    \label{fig:fig3}
\end{figure}

To evaluate the correlation among the elements of the activity time series $Y$, we investigate $\frac{Var[Y_{\Delta t}]}{Var[Y]}$, the ratio of the variance of the aggregated $Y_{\Delta t}$ to that of the original time series $Y$, as a function of $\Delta t$ (see Figure ~\ref{fig:fig3}). 

Firstly, we derive $\frac{\text{Var}[Y_{\Delta t}]}{\text{Var}[Y]}$ analytically for the general case and then prove that $\frac{\text{Var}[Y_{\Delta t}]}{\text{Var}[Y]}= \frac{1}{\Delta t}$, if each element of $Y$ is independent. 
The global activity $Y$ can be considered as a realization of a set of $T$ random variables $\{\hat{Y}(t)\}$, that are
identically distributed as a variable $\hat{Y}$. Hence, $Var[\hat{Y}(t)]=Var[\hat{Y}]$, where $t\in[0,T-1]$.
The variance of the aggregated activity $\hat{Y}_{\Delta t}(t')$ at a random time biin $t' $ with $0\leq t'\leq \left\lfloor\dfrac{T-1}{\Delta t}\right\rfloor -1$ follows

\begin{equation}
\begin{aligned}
    \text{Var}\bigg[\hat{Y}_{\Delta t}(t')\bigg] &=
    \text{Var}\bigg[\frac{1}{\Delta t}\sum_{t=t'\Delta t}^{t'\Delta t+\Delta t-1} \hat{Y}(t)\bigg]\\ &=  \frac{1}{(\Delta t) ^ 2}\sum_{t=t'\Delta t}^{t'\Delta t+\Delta t-1}\bigg( \text{Var}[\hat{Y}(t)] + 2\sum_{t'\Delta t\leq s < k\leq t'\Delta t+\Delta t-1} \text{Cov}[\hat{Y}(s),\hat{Y}(k)]\bigg)\\
    &=\frac{\text{Var}[\hat{Y}]}{\Delta t}+
    \frac{2}{(\Delta t) ^ 2}\sum_{t=t'\Delta t}^{t'\Delta t+\Delta t-1}\sum_{t'\Delta t\leq s < k\leq t'\Delta t+\Delta t-1} \text{Cov}[\hat{Y}(s),\hat{Y}(k)]
\end{aligned}
\label{eq:variance}
\end{equation}

When the activity $\hat{Y}(t)$ at each time $t$ is independently distributed, i.e. the set $\{\hat{Y}(t)\}$ are independent, the second term is zero and we have  $\frac{\text{Var}[\hat{Y}_{\Delta t}(t')]}{\text{Var}[\hat{Y}]}=\frac{1}{\Delta t}$. 

This explains why $\frac{\text{Var}[Y_{\Delta t}]}{\text{Var}[Y]}= \frac{1}{\Delta t}$ in Figure~\ref{fig:fig3} when we randomly re-shuffle the global activity $Y=\{y(t),t\in[0,T-1]\}$ in each of the thirteen temporal networks.
Figure~\ref{fig:fig3} shows that $\frac{\text{Var}[Y_{\Delta t}]}{\text{Var}[Y]} > \frac{1}{\Delta t}$ in all real-world temporal networks, suggesting the correlation among the number of contacts per time step at different time steps. Moreover, it is seen that the physical contact networks are further away from $\frac{\text{Var}[Y_{\Delta t}]}{\text{Var}[Y]}= \frac{1}{\Delta t}$ compared to the virtual contact networks, reflecting higher correlation in physical contacts than in virtual activities. The higher correlation in global activity in physical contacts networks than in virtual networks seems to be supported by the relatively higher probability for the global inter-event time to be relatively small in physical contact networks (see Figure \ref{fig:fig14}) than that in virtual contact ones (see Figure \ref{fig:fig13}).
\subsection{Topological and temporal distances between two contacts}
\label{sec:top_temp_analysis}
Next we wish to explore the relation between the topological distance and temporal distance of two contacts. Firstly, we explore whether contacts that are close in time are also close in topology. Contacts of temporal networks are measured at discrete time steps. The duration of each time step is either 1 or 20 seconds in the datasets listed in Table \ref{tab:1}. To compare physical and virtual contact networks, we present the time distance between any two contacts in units of seconds. 

\begin{figure}[H]

\includegraphics[width=1.\textwidth]{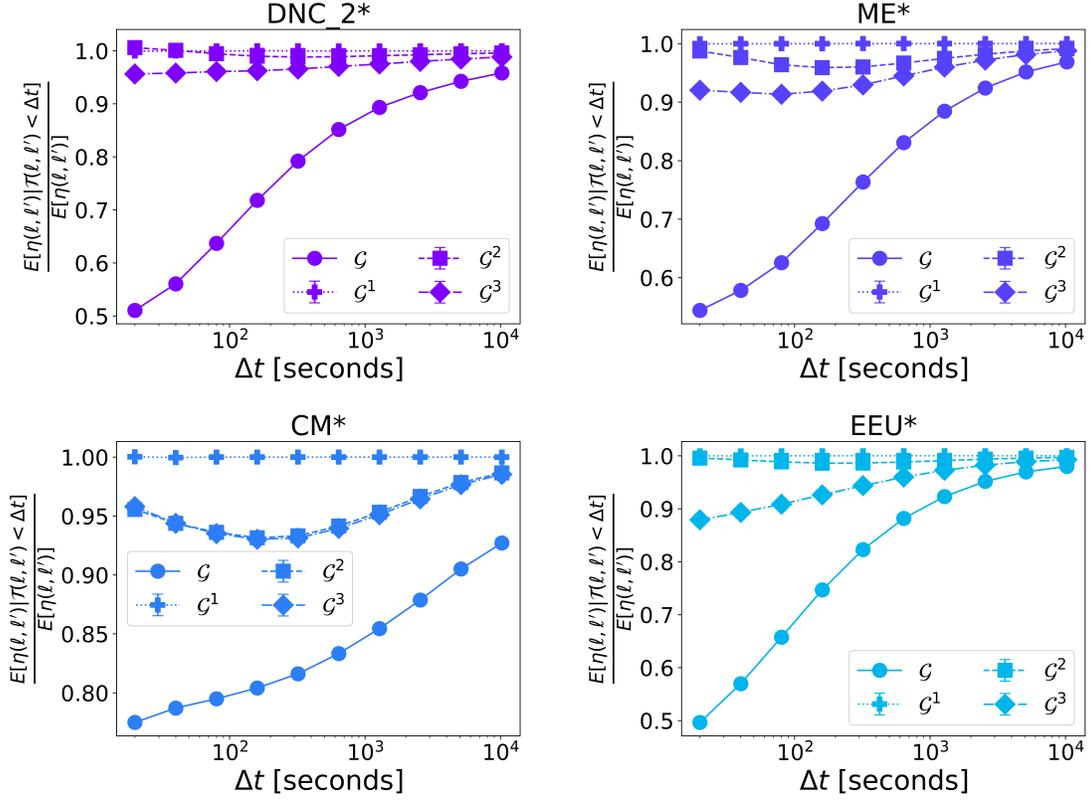}
\caption  {$\frac{E[\eta(\ell,\ell') | \mathcal{T} (\ell,\ell') < \Delta t ]}{E[\eta(\ell,\ell')]}$ as a function of $\Delta t$  for real world network $\mathcal{G} $ (points, solid line) and the three randomized reference models $\mathcal{G}^1$ (pluses, dotted line),  $\mathcal{G}^2$ (squares, dash line) and $\mathcal{G}^3$ (diamonds, dash-dotted line) in each virtual contact dataset. When $\frac{E[\eta(\ell,\ell') | \mathcal{T} (\ell,\ell') < \Delta t ]}{E[\mathcal{\eta}(\ell,\ell')]}=1$, topological and temporal distances are independent. Moreover, $\lim_{\Delta t\to\infty} E[\eta(\ell,\ell') | \mathcal{T} (\ell,\ell') < \Delta t ] = E[\eta(\ell,\ell')]$.
The results for each of the three randomized networks are obtained from 10 independent realizations of the randomized network.
Note that the horizontal  axis is presented in logarithmic scale.}
\label{fig:fig4}
\end{figure}

\begin{figure}[H]

\includegraphics[width=1\textwidth]{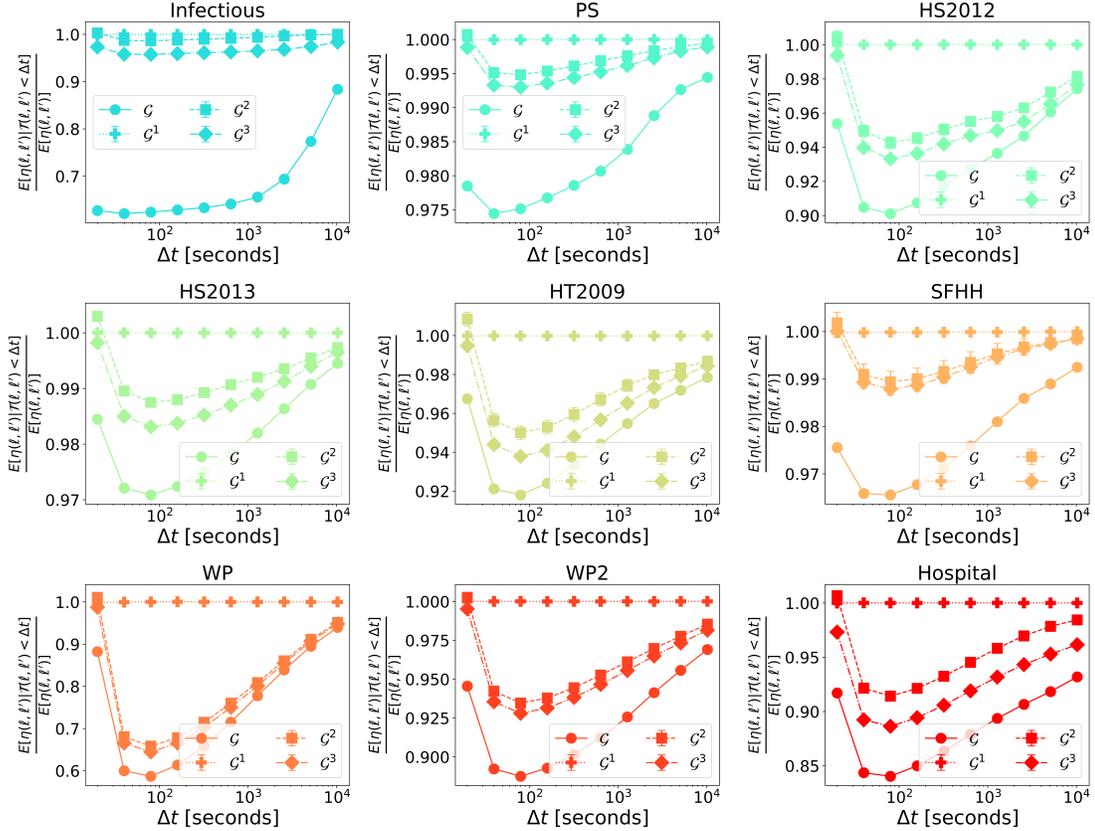}
\caption  {$\frac{E[\eta(\ell,\ell') | \mathcal{T} (\ell,\ell') < \Delta t ]}{E[\mathcal{\eta}(\ell,\ell')]}$ as a function of $\Delta t$ for real world network $\mathcal{G} $ (points, solid line) and the three randomized reference models $\mathcal{G}^1$ (pluses, dotted line),  $\mathcal{G}^2$ (squares, dash line) and $\mathcal{G}^3$ (diamonds, dash-dotted line) in each physical contact dataset. When $\frac{E[\eta(\ell,\ell') | \mathcal{T} (\ell,\ell') < \Delta t ]}{E[\mathcal{\eta}(\ell,\ell')]}=1$, topological and temporal distances are independent. Moreover, $\lim_{\Delta t\to\infty} E[\eta(\ell,\ell') | \mathcal{T} (\ell,\ell') < \Delta t ] = E[\eta(\ell,\ell')]$.
For each of the three randomized models, the lines and corresponding error bars correspond to the average and standard deviation of the results obtained from 10 independent realizations. Note that the horizontal  axis is presented in logarithmic scale.}
\label{fig:fig5}
\end{figure}

We analyze the average topological distance $E[\eta(\ell,\ell') | \mathcal{T} (\ell,\ell') < \Delta t ]$ of two contacts given that their temporal distance is less than $\Delta t$. If topological and temporal distances of two contacts are independent, $E[\eta(\ell,\ell') | \mathcal{T} (\ell,\ell') < \Delta t ] = E[\eta(\ell,\ell')]$ does not depend on the temporal distance $\Delta t$. Figures  \ref{fig:fig4} and \ref{fig:fig5} show that $E[\eta(\ell,\ell') | \mathcal{T} (\ell,\ell') < \Delta t ]$ increases with $\Delta t$ in real-world temporal networks. That is, contacts that are close in time are typically also close in topology. Such an increasing trend or correlation between temporal and topological distances in each real-world temporal network is evidently higher than that in the corresponding three randomized networks. Network $\mathcal{G}^{3}$ (swapping the activity time series of the two randomly selected links but with the same total number of contacts) preserves more properties of the original temporal network compared to $\mathcal{G}^{1}$ (swapping timestamps among contacts) and $\mathcal{G}^{2}$ (swapping the activity time series of two randomly selected links). Consistently, the increasing trend between temporal and topological distances 
is significantly reduced in $\mathcal{G}^{3}$, reduced further in $\mathcal{G}^{2}$ and disappears in $\mathcal{G}^{1}$. The slight initial decrease and afterwards increase of $E[\eta(\ell,\ell') | \mathcal{T} (\ell,\ell') < \Delta t ]$ with $\Delta t$ in $\mathcal{G}^{2}$  can be explained by the changes of the probability that a couple of contacts  with temporal distance smaller than $\Delta t$ are activations of the same link with $\Delta t$ (see the detailed discussion in Section \ref{sec:app_dist} of Appendix). The increase of $E[\eta(\ell,\ell') | \mathcal{T} (\ell,\ell') < \Delta t ]$ with $ \Delta t$ in $\mathcal{G}$, in comparison with that in $\mathcal{G}^{2}$, is more significant in  virtual contact networks and physical contact networks Infectious. In these networks, contacts that occur close in time tend to be close in topology. The stronger correlation in virtual networks and the physical network Infectious has also been observed when the other methods are used to characterize the temporal-topological correlation of contacts  (see Subsections \ref{subsub:temp_corr_ego} and \ref{sec:ego_vs_link}). The high correlation in network Infectious is related to the specific properties of this network. Network Infectious records the contacts among visitors of a museum and only people that visit the museum at a similar time could have contacts \citep{isella2011s}.  

We could also identify the temporal and topological correlation of contacts via $E[\mathcal{T}(\ell,\ell')|\eta(\ell,\ell') = j]$, the average temporal distance of two contacts given that their topological distance is $j$. However, this measure could be limited in distinguishing the difference among networks due to the
small diameter, i.e., the maximal hopcount of real-world networks. 

\subsection{Local events}
\label{sub:temp_corr_local}

In this section, we explore the temporal correlation of contacts that happen within (at any link of) a local neighborhood in the aggregated network. The local neighborhood refers to the ego-network $ego(e(i,j))$ centered at a link $e(i,j)$ which consists of the link itself and all its neighboring links that share a common node with the link $e(i,j)$. The objective is to understand whether and how local events are correlated in time, in forming trains (bursts) of events, where events within a burst have short inter-event times and trains are separated by a long inactive period. 

\subsubsection{Temporal correlation of contacts at an ego network}
\label{subsub:temp_corr_ego}

\begin{figure}[!h]
	\centering
	\includegraphics[width = 0.9\textwidth]{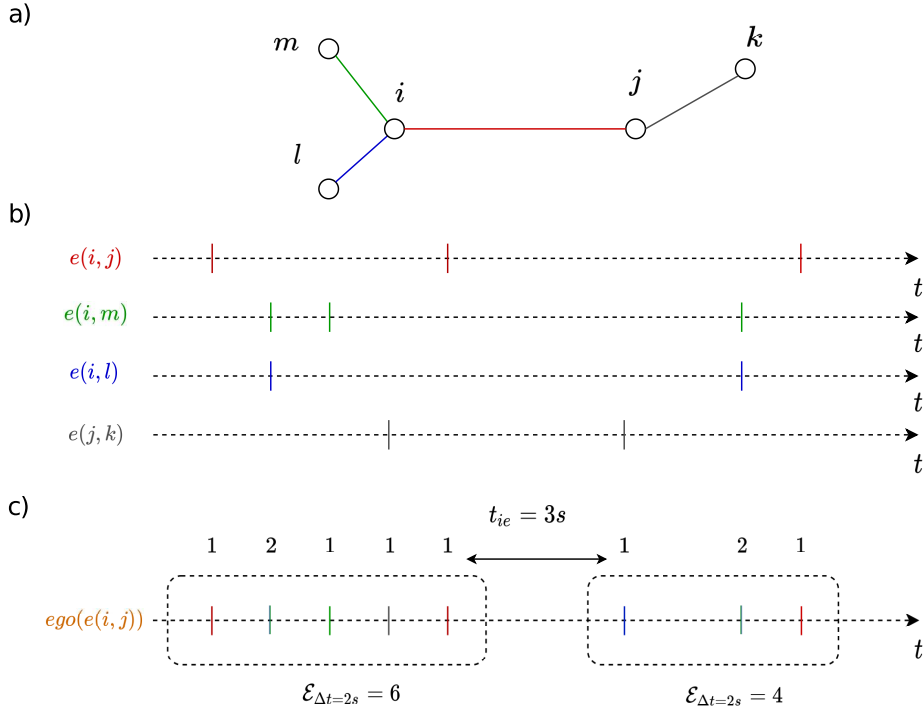}
	\caption{Schematic representation of  a) the ego-network of the link $e(i,j) $, i.e. $ego(e(i,j))$, b) the time series associated to each links in $ego(e(i,j))$ , c) the activity time series of  $ego(e(i,j))$, which is the sum of the time series of links belonging to the ego-network, and its event trains when $\Delta t = 2s$. }
	\label{fig6}
\end{figure}

We will analyze the activity (sequence) of an $ego(e(i,j))$ which records the number of contacts that happen within the ego-network at each time step, and equals the sum of the activity time series of every link in $ego(e(i,j))$ (see Figure \ref{fig6}). 

To evaluate correlation of local events in forming trains of events,  we study the train size distribution \citep{karsai2012correlated} of the activity sequence of an ego-network. A train of events is a sequence of consecutive contacts/events whose inter-event times are shorter than or equal to $\Delta t$ and separated from the other contacts by an inter-event time larger than $\Delta t$. Given a $\Delta t$, trains can be identified for each ego-network activity sequence (see Figure \ref{fig6}). Given a $\Delta t$
and a temporal network, the train size distribution $Pr[\mathcal{E}_{\Delta t} = s]$, i.e., the  probability that the size $\mathcal{E}_{\Delta t}$ of a random train is $s$ can be derived from the trains of all the ego-networks centered at every link. The train size distribution is compared between each real-world network and its three randomized networks. 

\begin{figure}[H]
    \centering
    \includegraphics[width = 0.9\textwidth]{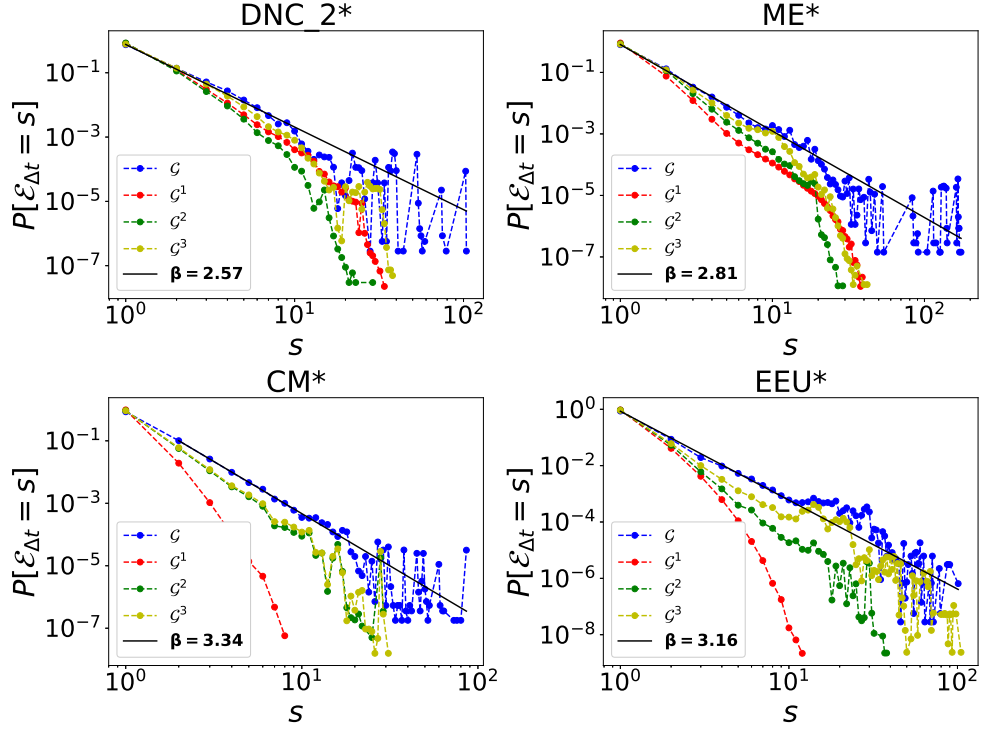}
    \caption{Train size distribution ($\Delta t = 60s$) of ego network activity for $\mathcal{G}$ (blue),  $\mathcal{G}_1$ (red),  $\mathcal{G}_2$ (green),  $\mathcal{G}_3$  (yellow) of virtual contact datasets. The black solid line represents the fit $P[\mathcal{E}_{\Delta t} = s] \sim s^{-\beta}$  to the distribution of the train size of $\mathcal{G}$ with $\Delta t =60s$. The power law fit and its fitting region were computed with Clauset's method  \citep{clauset2009power}. If the goodness of the power-law fit is significantly better than the exponential fit (likelihood ratio test with p-value $p<0.05$), the value of $\beta$ is reported in bold characters.}
    \label{fig:fig7}
\end{figure}

\begin{figure}[H]
    \centering
    \includegraphics[width = 0.9\textwidth]{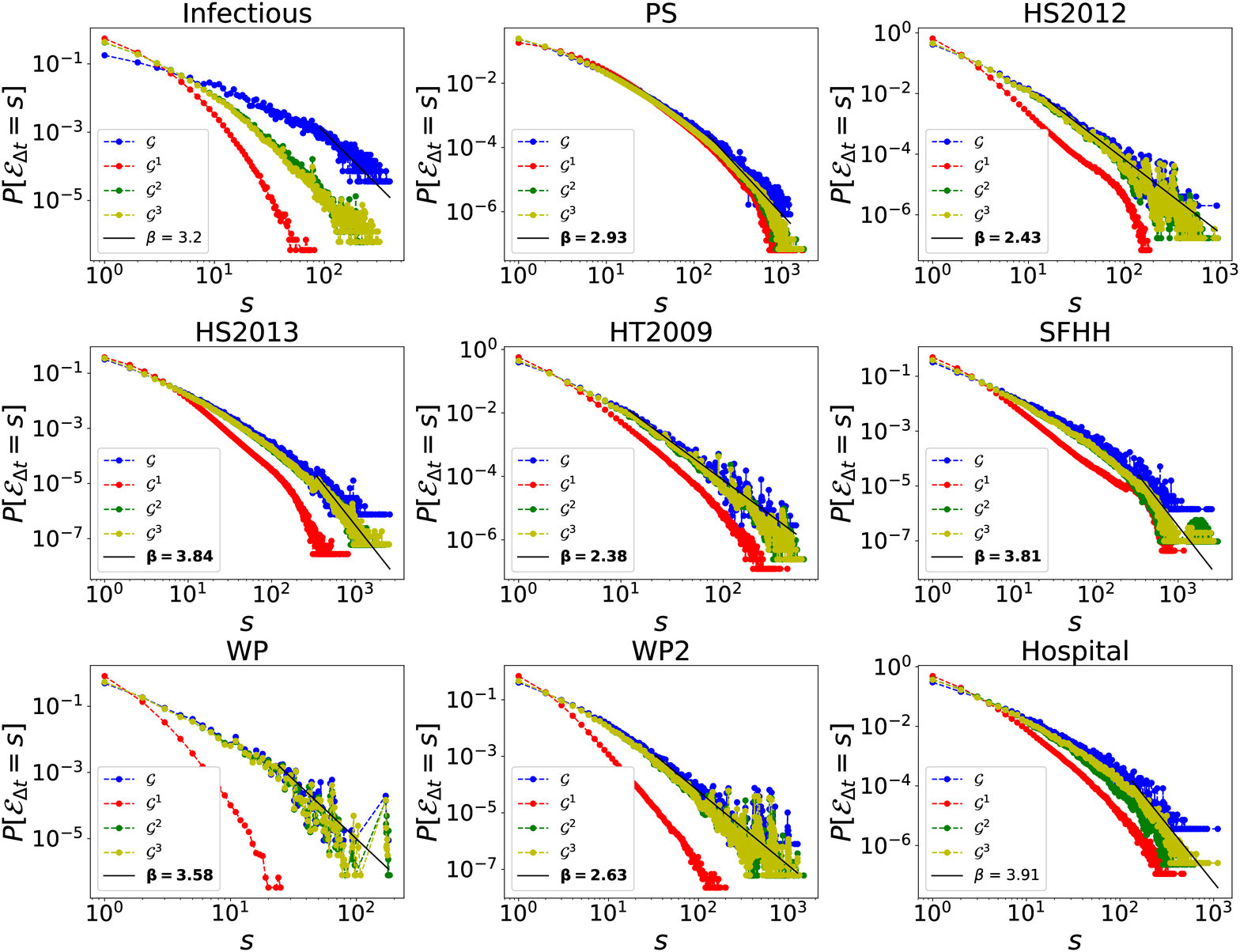}
    \caption{Train size distribution ($\Delta t = 60s$) of ego network activity for $\mathcal{G}$ (blue),  $\mathcal{G}_1$ (red),  $\mathcal{G}_2$ (green),  $\mathcal{G}_3$  (yellow) of physical contact datasets. The black solid line represents the fit $P[\mathcal{E}_{\Delta t} = s] \sim s^{-\beta}$  to the distribution of the train size of $\mathcal{G}$ with $\Delta t =60s$. The power law fit and its fitting region were computed with Clauset's method  \citep{clauset2009power}. If the goodness of the power-law fit is significantly better than the exponential fit (likelihood ratio test with p-value $p<0.05$), the value of $\beta$ is reported in bold characters.}
    \label{fig8}
\end{figure}

We find that the train size distribution when $\Delta t=60s$ in a real-world network has an evidently higher tail than that of the corresponding randomized networks in the four real-world virtual contact network (Figure  \ref{fig:fig7}) and the physical contact network Infectious (Figure  \ref{fig8}). Later, we will explain why $\Delta t=60s$ is representative. In these real-world networks, local events have a higher chance to form long trains, than in their corresponding randomized networks. 
Randomized network $\mathcal{G}_2$ is obtained by shuffling the activity sequences among the links, thus preserving the set of link activity sequences but removing their correlation with the network topology. The difference between the train size distribution in the ego-networks of $\mathcal{G}_2$ and an exponential distribution (the train size distribution when the inter-event times in the activity sequence are independent\footnote{The train size distribution follows an exponential function $Pr[\mathcal{E}_{\Delta t} = s]=Pr[t_{ie} \leq \Delta t]^{(s-1)}(1-Pr[t_{ie} \leq \Delta t])$ when the inter-event times in the activity sequence are independent.  Such exponential function is the product of the probability of observing s-1 inter-event times shorter than or equal to $\Delta t$, and a single inter-event time longer than $\Delta t$.}) reflects solely the temporal correlation of events in a link activity sequence in real-world networks. The different train size distributions in real-world networks $\mathcal{G}$ and their corresponding randomized networks $\mathcal{G}_2$ in Figures \ref{fig:fig7} and \ref{fig8} indicate that temporal correlation of activities at each link is insufficient to explain the temporal correlation of contacts at ego-networks. Instead, the correlation between the activity sequences and topology also contributes. Such temporal correlation of local activities suggests that neighboring nodes tend to have contacts or activities within a short time. The evidently stronger correlation observed in virtual networks and the physical network Infectious is in line with the finding in Section \ref{sec:top_temp_analysis}. 

The choice of $\Delta t$
is non-trivial. 
Karsai et al. \citep{karsai2012correlated} have observed a power-law train size distribution $Pr[\mathcal{E}_{\Delta t} = s] \sim s^{-\beta}$ of the activity of a link with $\beta = 0.39$ (0.42) in voice calls (SMS) temporal contact network and found that the power-law exponent remains approximately the same when $\Delta t$ varies within a broad range.
Our comparison of train size distribution with different $\Delta t$ for virtual (Figure \ref{fig:fig17}) and physical contact datasets (Figure \ref{fig:fig18}) shows that when $\Delta t$ is small ($\Delta t \leq 120 s$), the distribution is fat-tailed. The exponent $\beta$ of the power-law fit seems more stable across different values of $\Delta t$ in virtual contacts (Figure\ref{fig:fig17}) datasets than in physical contact ones (Figure \ref{fig:fig18}). The changes in the shape of the train size distribution of physical contact datasets are likely due to finite size effects which emerge because of limited duration of empirical temporal networks' observation window. When $\Delta t$ is sufficiently large, for example, any ego network has a single train, whose size is the total number of contacts that occur within the ego network. 
Figures \ref{fig:fig4} and \ref{fig:fig5} show that the positive correlation between topological and time distances (in linear scale) of two contacts is more evident when the time distance is small. Moreover, the observation time windows of temporal networks, especially physical contact networks, are short in duration. All these perspectives motivate us to consider a small $\Delta t$, e.g. $\Delta t=60s$. Moreover, our observations are similar when $\Delta t=120s$ and when $\Delta t=60s$ for all the analysis. Hence, we focus our discussion on $\Delta t=60s$ and all the results when $\Delta t=120s$ are given in the Appendix.

\subsubsection{Ego network activity versus link activity}
\label{sec:ego_vs_link}
We investigate further whether the temporal correlation of contacts that occur within an ego-network in forming long event trains could be explained or introduced by the temporal correlation of contacts at each single link. 

\begin{figure}[H]
\centering
\includegraphics[scale = 0.35]{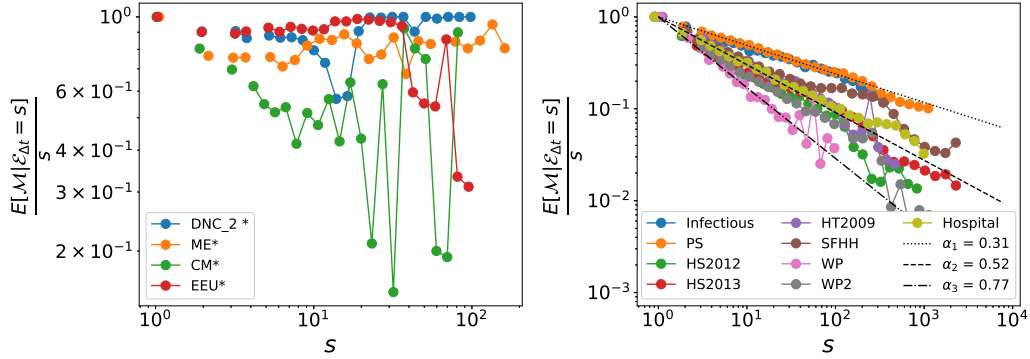}
\caption  {The average number of active links $\frac{E[\mathcal{M}|\mathcal{E}_{\Delta t}=s]}{s}$ for trains with size $\mathcal{E}_{\Delta t}=s$, normalized by the train size $s$ of the ego networks for virtual (left) and physical (right) contact datasets, when $\Delta t = 60s$. The three reference lines in right plot indicate $\frac{E[\mathcal{M}|\mathcal{E}_{\Delta t}=s]}{s} = s^{-\alpha}$ with slope $\alpha_1 = 0.31$ (dotted), $\alpha_2 = 0.52$ (dashed) and $\alpha_3 = 0.77$ (dash-dot). Note that the horizontal and vertical axes are presented in logarithmic scales. In total 30 logarithmic bins are split within the interval $[1, s_{max}]$, where $s_{max}$ is the largest train size observed in the considered real temporal network.} 
\label{fig:fig9}
\end{figure}

Firstly, we explore whether each activity train of an ego network contains the activities (contacts) of a single link or multiple links in the ego network. We examine the number $\mathcal{M}$ of distinct active links that a train of an ego-network involves. Specifically, each identified train of an ego network is composed of a set of contacts, occurring at a subset of links within the ego-network, the so-called active links. For each real-world network and given $\Delta t=60s$, trains are identified for every ego network centered at each link, and the number $\mathcal{M}$ of distinct active links of each train is counted. Figure \ref{fig:fig9} illustrates the average number of active links $\frac{E[\mathcal{M}|\mathcal{E}_{\Delta t}=s]}{s}$ for trains with size $\mathcal{E}_{\Delta t}=s$, normalized by the train size $s$, for virtual and physical contact networks, respectively. In all networks the fraction of active links $\frac{E[\mathcal{M}|\mathcal{E}_{\Delta t}=s]}{s}$ is above $\frac{E[\mathcal{M}|\mathcal{E}_{\Delta t}=s]}{s}=1/s$ suggesting that a train usually involves far more than $1$ active link. 
Interestingly, we observe in all 9 physical contact networks a seemingly power-law decay $\frac{E[\mathcal{M}|\mathcal{E}_{\Delta t}=s]}{s} \sim s^{-\alpha}$ (right plot of Figure \ref{fig:fig9}). In contrast, $\alpha \approx 0$, or equivalently $E[\mathcal{M}|\mathcal{E}_{\Delta t}=s] \sim s$ in virtual contact networks, especially mail dataset, i.e. EEU, ME and DNC2 (left plot of Figure \ref{fig:fig9}). This suggests that, in virtual contact networks, each train is mostly composed of the activities of many links in an ego network.

We compare further the physical contact networks. Their power-law exponents are within $0.31\leq\alpha\leq 0.77$. The slope of the power-law decay seems to be influenced by the type of human interaction and spacial constraints of the contact environment. Networks that lead to the slowest decay, i.e. $\alpha \approx 0.31$ are Infectious and PS datasets, which are contact networks in a museum and primary school respectively. The two contact networks of employees at a work place, WP and WP2 have the largest slope $\alpha \approx 0.77$. The other networks, i.e., contacts of high school students, conference participants have a power-law exponent in between $0.31\leq\alpha\leq 0.77$. A similar trend has been observed when $\Delta t = 120s$ (see Figure \ref{fig:fig23} in Appendix). These observations could be explained by the spatial constraints of contacts and the nature that younger students tend to interact with many others in an active period.
The bursty events of a train tend to engage the largest number of links in an ego network in network Infectious and PS than the other physical contact networks.  This could be due to the freedom for individuals to move in the museum and in the primary school (relative to the small museum/class room) and the tendency that primary school students interact with many others in an active period. The other way round, employees at a work place are confined in space (their offices) and tend to interact with limited number of colleagues during a train of activities. In this sense, virtual contacts are the least confined to space, leading thus to a larger number of active links than physical contacts.

\begin{figure}[H]
\centering
\includegraphics[width = 0.9\textwidth]{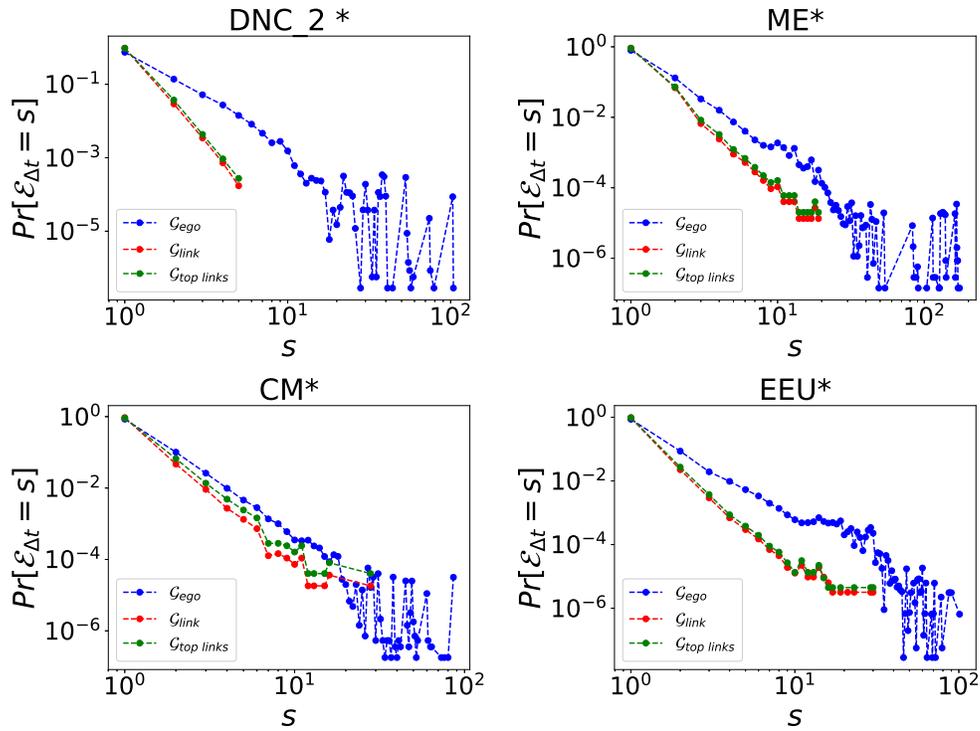}
\caption{Train size distribution ($\Delta t = 60s$) of ego network activity (blue),  single link activity (red),  most active link activity (green)  of virtual contact datasets. Note that the horizontal and vertical axes are presented in logarithmic scales.}
\label{fig:fig10}
\end{figure}

\begin{figure}[H]
\centering
\includegraphics[width = 0.9\textwidth]{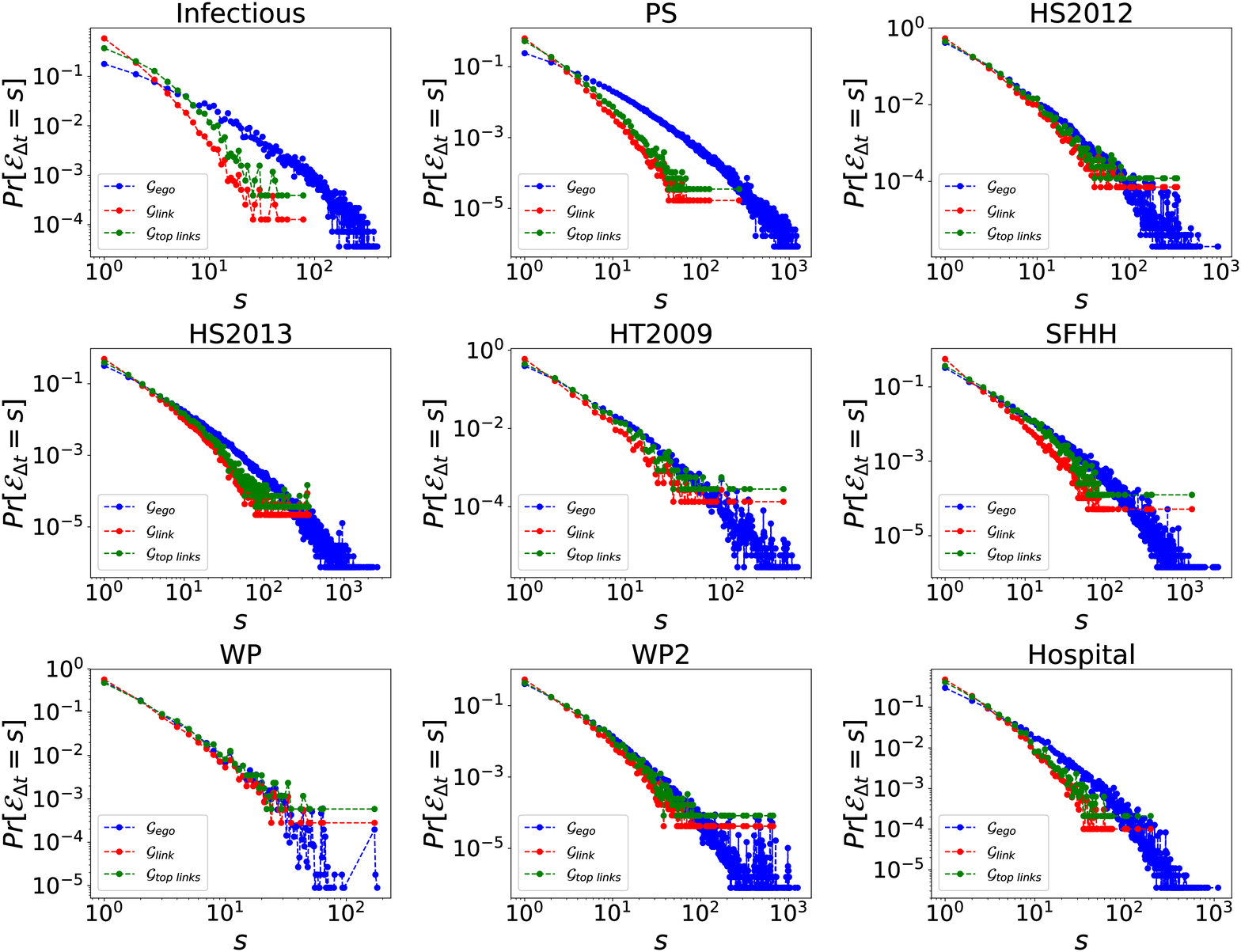}
\caption{Train size distribution ($\Delta t = 60s$) of ego-network activity (blue),  single link activity (red),  most active link activity (green)  of physical contact datasets. Note that the horizontal and vertical axes are presented in logarithmic scales.}
\label{fig:fig11}
\end{figure}

Whether each activity train of an ego network contains the activities of a single link or of multiple links could also be reflected via the train size distribution in an ego network versus the train size distribution in a link. In Figures \ref{fig:fig10} and \ref{fig:fig11}, we compare the train size distribution (with $\Delta t=60s$) of the activity sequence of single links, of the most active single links (top $10 \%$ of links with the largest number of contacts) and of ego-networks. The trains of ego-networks tend to be longer than those of single links and the most active single links, in all networks except for WP. Therefore, the trains of the ego network are usually the results of the activity of more than one link. The same observations are obtained when $\Delta t = 120s$ (see Figures \ref{fig:fig21} and \ref{fig:fig22} in Appendix). The similar train size distribution in ego networks and in links in the WP dataset is consistent with the largest power-exponent observed in Figure \ref{fig:fig9}. In WP, a train of an ego network is composed of the activity of relatively few links.

\subsubsection{Ego network activity versus node activity}

We further address the question whether a train at an ego network $ego(e(i,j))$ involves the activity of both end nodes $i$ and $j$, or only one of them. Event trains at ego networks engaging activities of both end nodes may suggest a possible social contagion in activity between nodes.

For each train of an ego network $ego(e(i,j))$, we consider the events that associate with only one end node but not both. Among these events, we count the fraction of events $\phi_i$ and $\phi_j$ that associate with end node $i$ and $j$ respectively and $\phi_i+\phi_j=1$. The maximum of the two fractions $B= max(\phi_i,\phi_i)$ quantifies how unbalanced the activities of the two end nodes $i$ and $j$ contribute to a train and is called the activity balance of a train.

Table~\ref{tab:activity_balance} shows the average activity balance $E[B]$ and the probability $Pr[B\leq 0.95]$ of the activity balance for all contact networks, accounting all trains (whose sizes are larger than 1) of all ego networks. We find that the average activity $E[B]<1$, suggesting that an activity train in an ego network $ego(e(i,j))$ engages in the activity of both end nodes $i$ and $j$. This is in line with the previous finding that the activity correlation in an ego network cannot be explained by the activity of a single link. Moreover, the activity is found to be larger, thus more unbalanced, in virtual contacts than in physical contact networks. This is likely because, in a virtual contact network like email contact network, an individual tends to contact many others at a similar time. A train of events at an ego network $ego(e(i,j))$ in a virtual network contains mainly the activities of a single end node $i$ or $j$. 


\begin{table}[H]
\begin{adjustbox}{width=\columnwidth,center}
\begin{tabular}{@{}llllllr@{}}
\toprule

& \multicolumn{3}{c}{\textbf{Virtual Contacts}} & \multicolumn{3}{c}{\textbf{Physical Contacts}}\\
\hline
& Dataset & $P[B\leq 0.95]$ & $E[B]$
    & Dataset & $P[B\leq 0.95]$ & $E[B]$  \\
    \midrule
&DNC 2* &0.05 &0.98 &Infectious &0.38 &0.87\\
&ME* &0.12 &0.95 &PS &0.44 &0.86 \\
&CM* &0.03 &0.99 &HS2012 &0.16 &0.95\\
&EEU* &0.12 &0.94 &HS2013 &0.25 &0.93 \\
& & & &HT2009 &0.21 &0.94 \\
& & & &SFHH &0.19 &0.95 \\
& & & &WP &0.06 &0.98 \\
& & & &WP2 &0.1 &0.97 \\
& & & &Hospital &0.12 &0.97\\

\end{tabular}
\end{adjustbox}
\caption{Probability $P[B\leq 0.95]$ and average $E[B]$ of the activity balance $B$ in virtual contact (left) and physical contact (right) networks.}
\label{tab:activity_balance}
\end{table}

\section{Conclusions}

In this paper, we developed systematically methods to characterize jointly the topological and temporal properties of contacts in a time-evolving network, ranging from global network level to local neighborhoods. Via applying these methods to real-world networks, we identified substantial differences between virtual and physical contact networks. 

We find that contacts that occur close in time tend to be close in topology and this trend is more evident in virtual contact networks compared to physical contact networks. This is in line with the observation that the contacts within an ego-networks tend to have a higher chance to form long trains and thus happen closely in time in real-world networks. Such activity correlation is more evident in virtual contact networks. Moreover, an event train of an ego network $ego(e(i,j))$ is mostly composed of the activities of multiple component links. Interestingly, more links tend to be engaged in e.g., virtual networks and physical contact network primary school where the contacts are less constrained in space, in contrast to e.g., the contact network at workplace. These may suggest that contacts with a low cost may better facilitate social contagion, i.e. influence between neighboring nodes in the activity. Finally, an event train of an ego network $ego(e(i,j))$ usually contains the activity of both ends, node $i$ and $j$. Two connected nodes, thus, tend to have contacts with their neighbors close in time. The two end nodes' contributions are more unbalanced in virtual contacts than in physical contacts, likely driven by the nature that in a virtual (e.g. email) contact network, an individual tends to contact many others close in time.

Our methods are confined to undirected networks. A full-fledged directed temporal network characterization method is deemed as promising to develop. The application of these methods may enhance our understanding of diverse time-evolving systems and allow exploration of the influence of detected properties/patterns on a dynamic process upon the network. Finally, the detected patterns may further inspire the development of more realistic temporal network models that reproduce key realistic temporal and topological properties of contacts.

\bibliographystyle{unsrtnat}
\bibliography{M335}

\begin{thebibliography}{48}
\providecommand{\natexlab}[1]{#1}
\providecommand{\url}[1]{\texttt{#1}}
\expandafter\ifx\csname urlstyle\endcsname\relax
  \providecommand{\doi}[1]{doi: #1}\else
  \providecommand{\doi}{doi: \begingroup \urlstyle{rm}\Url}\fi

\bibitem[Holme and Saram{\"a}ki(2012)]{holme2012temporal}
Petter Holme and Jari Saram{\"a}ki.
\newblock Temporal networks.
\newblock \emph{Physics Reports}, 519\penalty0 (3):\penalty0 97--125, 2012.

\bibitem[Holme(2015)]{holme2015modern}
Petter Holme.
\newblock Modern temporal network theory: a colloquium.
\newblock \emph{The European Physical Journal B}, 88\penalty0 (9):\penalty0
  234, 2015.

\bibitem[Goh and Barab{\'a}si(2008)]{goh2008burstiness}
K-I Goh and A-L Barab{\'a}si.
\newblock Burstiness and memory in complex systems.
\newblock \emph{EPL (Europhysics Letters)}, 81\penalty0 (4):\penalty0 48002,
  2008.

\bibitem[Eckmann et~al.(2004)Eckmann, Moses, and Sergi]{eckmann2004entropy}
Jean-Pierre Eckmann, Elisha Moses, and Danilo Sergi.
\newblock Entropy of dialogues creates coherent structures in e-mail traffic.
\newblock \emph{Proceedings of the National Academy of Sciences}, 101\penalty0
  (40):\penalty0 14333--14337, 2004.

\bibitem[Oliveira and Barab{\'a}si(2005)]{oliveira2005darwin}
Joao~Gama Oliveira and Albert-L{\'a}szl{\'o} Barab{\'a}si.
\newblock Darwin and einstein correspondence patterns.
\newblock \emph{Nature}, 437\penalty0 (7063):\penalty0 1251--1251, 2005.

\bibitem[Candia et~al.(2008)Candia, Gonz{\'a}lez, Wang, Schoenharl, Madey, and
  Barab{\'a}si]{candia2008uncovering}
Juli{\'a}n Candia, Marta~C Gonz{\'a}lez, Pu~Wang, Timothy Schoenharl, Greg
  Madey, and Albert-L{\'a}szl{\'o} Barab{\'a}si.
\newblock Uncovering individual and collective human dynamics from mobile phone
  records.
\newblock \emph{Journal of Physics A: Mathematical and Theoretical},
  41\penalty0 (22):\penalty0 224015, 2008.

\bibitem[Johansen(2004)]{johansen2004probing}
Anders Johansen.
\newblock Probing human response times.
\newblock \emph{Physica A: Statistical Mechanics and its Applications},
  338\penalty0 (1-2):\penalty0 286--291, 2004.

\bibitem[Barab{\'a}si and Bonabeau(2003)]{barabasi2003scale}
Albert-L{\'a}szl{\'o} Barab{\'a}si and Eric Bonabeau.
\newblock Scale-free networks.
\newblock \emph{Scientific American}, 288\penalty0 (5):\penalty0 60--69, 2003.

\bibitem[Barab{\'a}si(2009)]{barabasi2009scale}
Albert-L{\'a}szl{\'o} Barab{\'a}si.
\newblock Scale-free networks: a decade and beyond.
\newblock \emph{Science}, 325\penalty0 (5939):\penalty0 412--413, 2009.

\bibitem[Newman(2003)]{newman2003structure}
Mark~EJ Newman.
\newblock The structure and function of complex networks.
\newblock \emph{SIAM review}, 45\penalty0 (2):\penalty0 167--256, 2003.

\bibitem[Boccaletti et~al.(2006)Boccaletti, Latora, Moreno, Chavez, and
  Hwang]{boccaletti2006complex}
Stefano Boccaletti, Vito Latora, Yamir Moreno, Martin Chavez, and D-U Hwang.
\newblock Complex networks: Structure and dynamics.
\newblock \emph{Physics Reports}, 424\penalty0 (4-5):\penalty0 175--308, 2006.

\bibitem[Barrat et~al.(2004)Barrat, Barthelemy, Pastor-Satorras, and
  Vespignani]{barrat2004architecture}
Alain Barrat, Marc Barthelemy, Romualdo Pastor-Satorras, and Alessandro
  Vespignani.
\newblock The architecture of complex weighted networks.
\newblock \emph{Proceedings of the National Academy of Sciences}, 101\penalty0
  (11):\penalty0 3747--3752, 2004.

\bibitem[Onnela et~al.(2007)Onnela, Saram{\"a}ki, Hyv{\"o}nen, Szab{\'o},
  Lazer, Kaski, Kert{\'e}sz, and Barab{\'a}si]{onnela2007structure}
J-P Onnela, Jari Saram{\"a}ki, Jorkki Hyv{\"o}nen, Gy{\"o}rgy Szab{\'o}, David
  Lazer, Kimmo Kaski, J{\'a}nos Kert{\'e}sz, and A-L Barab{\'a}si.
\newblock Structure and tie strengths in mobile communication networks.
\newblock \emph{Proceedings of the National Academy of Sciences}, 104\penalty0
  (18):\penalty0 7332--7336, 2007.

\bibitem[Brot et~al.(2016)Brot, Muchnik, Goldenberg, and
  Louzoun]{brot2016evolution}
Hilla Brot, Lev Muchnik, Jacob Goldenberg, and Yoram Louzoun.
\newblock Evolution through bursts: Network structure develops through
  localized bursts in time and space.
\newblock \emph{Network Science}, 4\penalty0 (3):\penalty0 293--313, 2016.

\bibitem[Kikas et~al.(2013)Kikas, Dumas, and Karsai]{kikas2013bursty}
Riivo Kikas, Marlon Dumas, and M{\'a}rton Karsai.
\newblock Bursty egocentric network evolution in skype.
\newblock \emph{Social Network Analysis and Mining}, 3\penalty0 (4):\penalty0
  1393--1401, 2013.

\bibitem[Paranjape et~al.(2017)Paranjape, Benson, and
  Leskovec]{paranjape2017motifs}
Ashwin Paranjape, Austin~R Benson, and Jure Leskovec.
\newblock Motifs in temporal networks.
\newblock In \emph{Proceedings of the Tenth ACM International Conference on Web
  Search and Data Mining}, pages 601--610, 2017.

\bibitem[Kovanen et~al.(2011)Kovanen, Karsai, Kaski, Kert{\'e}sz, and
  Saram{\"a}ki]{kovanen2011temporal}
Lauri Kovanen, M{\'a}rton Karsai, Kimmo Kaski, J{\'a}nos Kert{\'e}sz, and Jari
  Saram{\"a}ki.
\newblock Temporal motifs in time-dependent networks.
\newblock \emph{Journal of Statistical Mechanics: Theory and Experiment},
  2011\penalty0 (11):\penalty0 P11005, 2011.

\bibitem[Karsai et~al.(2012{\natexlab{a}})Karsai, Kaski, and
  Kert{\'e}sz]{karsai2012correlated}
M{\'a}rton Karsai, Kimmo Kaski, and J{\'a}nos Kert{\'e}sz.
\newblock Correlated dynamics in egocentric communication networks.
\newblock \emph{Plos One}, 7\penalty0 (7):\penalty0 e40612, 2012{\natexlab{a}}.

\bibitem[Pan and Saram{\"a}ki(2011)]{pan2011path}
Raj~Kumar Pan and Jari Saram{\"a}ki.
\newblock Path lengths, correlations, and centrality in temporal networks.
\newblock \emph{Physical Review E}, 84\penalty0 (1):\penalty0 016105, 2011.

\bibitem[Barabasi(2005)]{barabasi2005origin}
Albert-Laszlo Barabasi.
\newblock The origin of bursts and heavy tails in human dynamics.
\newblock \emph{Nature}, 435\penalty0 (7039):\penalty0 207--211, 2005.

\bibitem[Vazquez et~al.(2007)Vazquez, Racz, Lukacs, and
  Barabasi]{vazquez2007impact}
Alexei Vazquez, Balazs Racz, Andras Lukacs, and Albert-Laszlo Barabasi.
\newblock Impact of non-poissonian activity patterns on spreading processes.
\newblock \emph{Physical Review Letters}, 98\penalty0 (15):\penalty0 158702,
  2007.

\bibitem[Rybski et~al.(2009)Rybski, Buldyrev, Havlin, Liljeros, and
  Makse]{rybski2009scaling}
Diego Rybski, Sergey~V Buldyrev, Shlomo Havlin, Fredrik Liljeros, and
  Hern{\'a}n~A Makse.
\newblock Scaling laws of human interaction activity.
\newblock \emph{Proceedings of the National Academy of Sciences}, 106\penalty0
  (31):\penalty0 12640--12645, 2009.

\bibitem[Karsai et~al.(2012{\natexlab{b}})Karsai, Kaski, Barab{\'a}si, and
  Kert{\'e}sz]{karsai2012universal}
M{\'a}rton Karsai, Kimmo Kaski, Albert-L{\'a}szl{\'o} Barab{\'a}si, and
  J{\'a}nos Kert{\'e}sz.
\newblock Universal features of correlated bursty behaviour.
\newblock \emph{Scientific Reports}, 2:\penalty0 397, 2012{\natexlab{b}}.

\bibitem[Zhan et~al.(2019{\natexlab{a}})Zhan, Hanjalic, and
  Wang]{zhan2019information}
Xiu-Xiu Zhan, Alan Hanjalic, and Huijuan Wang.
\newblock Information diffusion backbones in temporal networks.
\newblock \emph{Scientific Reports}, 9\penalty0 (1):\penalty0 1--12,
  2019{\natexlab{a}}.

\bibitem[Zhan et~al.(2019{\natexlab{b}})Zhan, Hanjalic, and
  Wang]{zhan2019suppressing}
Xiu-Xiu Zhan, Alan Hanjalic, and Huijuan Wang.
\newblock Suppressing information diffusion via link blocking in temporal
  networks.
\newblock In \emph{International Conference on Complex Networks and Their
  Applications}, pages 448--458. Springer, 2019{\natexlab{b}}.

\bibitem[Pfitzner et~al.(2013)Pfitzner, Scholtes, Garas, Tessone, and
  Schweitzer]{pfitzner2013betweenness}
Ren{\'e} Pfitzner, Ingo Scholtes, Antonios Garas, Claudio~J Tessone, and Frank
  Schweitzer.
\newblock Betweenness preference: Quantifying correlations in the topological
  dynamics of temporal networks.
\newblock \emph{Physical Review Letters}, 110\penalty0 (19):\penalty0 198701,
  2013.

\bibitem[Miritello et~al.(2011)Miritello, Moro, and
  Lara]{miritello2011dynamical}
Giovanna Miritello, Esteban Moro, and Rub{\'e}n Lara.
\newblock Dynamical strength of social ties in information spreading.
\newblock \emph{Physical Review E}, 83\penalty0 (4):\penalty0 045102, 2011.

\bibitem[Kivel{\"a} et~al.(2012)Kivel{\"a}, Pan, Kaski, Kert{\'e}sz,
  Saram{\"a}ki, and Karsai]{kivela2012multiscale}
Mikko Kivel{\"a}, Raj~Kumar Pan, Kimmo Kaski, J{\'a}nos Kert{\'e}sz, Jari
  Saram{\"a}ki, and M{\'a}rton Karsai.
\newblock Multiscale analysis of spreading in a large communication network.
\newblock \emph{Journal of Statistical Mechanics: Theory and Experiment},
  2012\penalty0 (03):\penalty0 P03005, 2012.

\bibitem[Scholtes et~al.(2014)Scholtes, Wider, Pfitzner, Garas, Tessone, and
  Schweitzer]{scholtes2014causality}
Ingo Scholtes, Nicolas Wider, Ren{\'e} Pfitzner, Antonios Garas, Claudio~J
  Tessone, and Frank Schweitzer.
\newblock Causality-driven slow-down and speed-up of diffusion in non-markovian
  temporal networks.
\newblock \emph{Nature Communications}, 5\penalty0 (1):\penalty0 1--9, 2014.

\bibitem[Williams et~al.(2019)Williams, Lillo, and Latora]{williams2019auto}
Oliver~E Williams, Fabrizio Lillo, and Vito Latora.
\newblock How auto-and cross-correlations in link dynamics influence diffusion
  in non-markovian temporal networks.
\newblock \emph{arXiv preprint arXiv:1909.08134}, 2019.

\bibitem[Backlund et~al.(2014)Backlund, Saram{\"a}ki, and
  Pan]{backlund2014effects}
Ville-Pekka Backlund, Jari Saram{\"a}ki, and Raj~Kumar Pan.
\newblock Effects of temporal correlations on cascades: Threshold models on
  temporal networks.
\newblock \emph{Physical Review E}, 89\penalty0 (6):\penalty0 062815, 2014.

\bibitem[Parshani et~al.(2010)Parshani, Dickison, Cohen, Stanley, and
  Havlin]{parshani2010dynamic}
Roni Parshani, Mark Dickison, Reuven Cohen, H~Eugene Stanley, and Shlomo
  Havlin.
\newblock Dynamic networks and directed percolation.
\newblock \emph{EPL (Europhysics Letters)}, 90\penalty0 (3):\penalty0 38004,
  2010.

\bibitem[Horv{\'a}th and Kert{\'e}sz(2014)]{horvath2014spreading}
D{\'a}vid~X Horv{\'a}th and J{\'a}nos Kert{\'e}sz.
\newblock Spreading dynamics on networks: the role of burstiness, topology and
  non-stationarity.
\newblock \emph{New Journal of Physics}, 16\penalty0 (7):\penalty0 073037,
  2014.

\bibitem[Delvenne et~al.(2015)Delvenne, Lambiotte, and
  Rocha]{delvenne2015diffusion}
Jean-Charles Delvenne, Renaud Lambiotte, and Luis~EC Rocha.
\newblock Diffusion on networked systems is a question of time or structure.
\newblock \emph{Nature Communications}, 6\penalty0 (1):\penalty0 1--10, 2015.

\bibitem[Kunegis(2013)]{kunegis2013konect}
J{\'e}r{\^o}me Kunegis.
\newblock Konect: the koblenz network collection.
\newblock In \emph{Proceedings of the 22nd international conference on World
  Wide Web}, pages 1343--1350, 2013.

\bibitem[Michalski et~al.(2011)Michalski, Palus, and Kazienko]{emailsRadoslaw}
Rados{\l}aw Michalski, Sebastian Palus, and Przemys{\l}aw Kazienko.
\newblock Matching organizational structure and social network extracted from
  email communication.
\newblock In \emph{Lecture Notes in Business Information Processing},
  volume~87, pages 197--206. Springer Berlin Heidelberg, 2011.

\bibitem[Panzarasa et~al.(2009)Panzarasa, Opsahl, and
  Carley]{panzarasa2009patterns}
Pietro Panzarasa, Tore Opsahl, and Kathleen~M Carley.
\newblock Patterns and dynamics of users' behavior and interaction: Network
  analysis of an online community.
\newblock \emph{Journal of the American Society for Information Science and
  Technology}, 60\penalty0 (5):\penalty0 911--932, 2009.

\bibitem[Leskovec et~al.(2007)Leskovec, Kleinberg, and
  Faloutsos]{leskovec2007graph}
Jure Leskovec, Jon Kleinberg, and Christos Faloutsos.
\newblock Graph evolution: Densification and shrinking diameters.
\newblock \emph{ACM transactions on Knowledge Discovery from Data (TKDD)},
  1\penalty0 (1):\penalty0 2--es, 2007.

\bibitem[Isella et~al.(2011)Isella, Stehl{\'e}, Barrat, Cattuto, Pinton, and
  Van~den Broeck]{isella2011s}
Lorenzo Isella, Juliette Stehl{\'e}, Alain Barrat, Ciro Cattuto,
  Jean-Fran{\c{c}}ois Pinton, and Wouter Van~den Broeck.
\newblock What's in a crowd? analysis of face-to-face behavioral networks.
\newblock \emph{Journal of theoretical biology}, 271\penalty0 (1):\penalty0
  166--180, 2011.

\bibitem[Stehl{\'e} et~al.(2011{\natexlab{a}})Stehl{\'e}, Voirin, Barrat,
  Cattuto, Isella, Pinton, Quaggiotto, Van~den Broeck, R{\'e}gis, Lina,
  et~al.]{stehle2011high}
Juliette Stehl{\'e}, Nicolas Voirin, Alain Barrat, Ciro Cattuto, Lorenzo
  Isella, Jean-Fran{\c{c}}ois Pinton, Marco Quaggiotto, Wouter Van~den Broeck,
  Corinne R{\'e}gis, Bruno Lina, et~al.
\newblock High-resolution measurements of face-to-face contact patterns in a
  primary school.
\newblock \emph{PloS one}, 6\penalty0 (8):\penalty0 e23176, 2011{\natexlab{a}}.

\bibitem[Fournet and Barrat(2014)]{fournet2014contact}
Julie Fournet and Alain Barrat.
\newblock Contact patterns among high school students.
\newblock \emph{PloS one}, 9\penalty0 (9):\penalty0 e107878, 2014.

\bibitem[Mastrandrea et~al.(2015)Mastrandrea, Fournet, and
  Barrat]{mastrandrea2015contact}
Rossana Mastrandrea, Julie Fournet, and Alain Barrat.
\newblock Contact patterns in a high school: a comparison between data
  collected using wearable sensors, contact diaries and friendship surveys.
\newblock \emph{PloS one}, 10\penalty0 (9):\penalty0 e0136497, 2015.

\bibitem[Cattuto et~al.(2010)Cattuto, Van~den Broeck, Barrat, Colizza, Pinton,
  and Vespignani]{cattuto2010dynamics}
Ciro Cattuto, Wouter Van~den Broeck, Alain Barrat, Vittoria Colizza,
  Jean-Fran{\c{c}}ois Pinton, and Alessandro Vespignani.
\newblock Dynamics of person-to-person interactions from distributed rfid
  sensor networks.
\newblock \emph{PloS one}, 5\penalty0 (7):\penalty0 e11596, 2010.

\bibitem[Stehl{\'e} et~al.(2011{\natexlab{b}})Stehl{\'e}, Voirin, Barrat,
  Cattuto, Colizza, Isella, R{\'e}gis, Pinton, Khanafer, Van~den Broeck,
  et~al.]{stehle2011simulation}
Juliette Stehl{\'e}, Nicolas Voirin, Alain Barrat, Ciro Cattuto, Vittoria
  Colizza, Lorenzo Isella, Corinne R{\'e}gis, Jean-Fran{\c{c}}ois Pinton,
  Nagham Khanafer, Wouter Van~den Broeck, et~al.
\newblock Simulation of an seir infectious disease model on the dynamic contact
  network of conference attendees.
\newblock \emph{BMC medicine}, 9\penalty0 (1):\penalty0 1--15,
  2011{\natexlab{b}}.

\bibitem[G{\'e}nois et~al.(2015)G{\'e}nois, Vestergaard, Fournet, Panisson,
  Bonmarin, and Barrat]{genois2015data}
Mathieu G{\'e}nois, Christian~L Vestergaard, Julie Fournet, Andr{\'e} Panisson,
  Isabelle Bonmarin, and Alain Barrat.
\newblock Data on face-to-face contacts in an office building suggest a
  low-cost vaccination strategy based on community linkers.
\newblock \emph{Network Science}, 3\penalty0 (3):\penalty0 326--347, 2015.

\bibitem[G{\'e}nois and Barrat(2018)]{genois2018can}
Mathieu G{\'e}nois and Alain Barrat.
\newblock Can co-location be used as a proxy for face-to-face contacts?
\newblock \emph{EPJ Data Science}, 7\penalty0 (1):\penalty0 1--18, 2018.

\bibitem[Vanhems et~al.(2013)Vanhems, Barrat, Cattuto, Pinton, Khanafer,
  R{\'e}gis, Kim, Comte, and Voirin]{vanhems2013estimating}
Philippe Vanhems, Alain Barrat, Ciro Cattuto, Jean-Fran{\c{c}}ois Pinton,
  Nagham Khanafer, Corinne R{\'e}gis, Byeul-a Kim, Brigitte Comte, and Nicolas
  Voirin.
\newblock Estimating potential infection transmission routes in hospital wards
  using wearable proximity sensors.
\newblock \emph{PloS one}, 8\penalty0 (9):\penalty0 e73970, 2013.

\bibitem[Clauset et~al.(2009)Clauset, Shalizi, and Newman]{clauset2009power}
Aaron Clauset, Cosma~Rohilla Shalizi, and Mark~EJ Newman.
\newblock Power-law distributions in empirical data.
\newblock \emph{SIAM review}, 51\penalty0 (4):\penalty0 661--703, 2009.

\end{thebibliography}

\appendix

\section{Appendix}

\subsection{Datasets Description (* indicates virtual contacts)}
\label{sec:dataset_descr}
\begin{itemize}
    \item 	\textbf{Manifacturing Email (ME) *}: Emails exchanged between 167 employees of a mid-size company in Poland, observation time: 270 days, time resolution 1 s
	\item \textbf{European Union Mail(EEU) *}: Emails exchanged between 986 accounts of a large European research institution during a period from October 2003 to May 2005 (18 months), time resolution 1 s
	\item \textbf{Democratic National Committee Mail *}: Emails of 1900 members (1598 after preprocessing) of the Democratic National Committee, in our case only final 33 days were considered, because they are more than 95\% of the entire corpus of email, time resolution 1 s
	\item \textbf{College Messages (CM)} *:  messages from an online community of 1899 (1892 after preprocessing) students at the University of California, Irvine. Time span of approximately 6 months, time resolution 1 s

	\item \textbf{Hypertext 2009 (HT09)} face-to-face interactions (Rfid sensors, range of 1.5-2 m, time resolution of 20s) of the 113 participants to Hypertext conference, during 3 days. 
	\item \textbf{Infectious (Science Gallery, Dublin)} face-to-face interactions (Rfid sensors, range of 1.5-2 m, time resolution of 20s) of 14000 visitors (410 after preprocessing) at the Science Gallery of Dublin, during 3 months of observation (after preprocessing, i.e. selecting the largest connected component, 1 day). Community structure linked to time of visit (only visitors present at the same time can interact)
	\item \textbf{Workplace (WP)} face-to-face interactions (Rfid sensors, range of 1.5-2 m, time resolution of 20s) of 92 employees in one of the two office buildings of the InVS, located in Saint Maurice near Paris, France, during two weeks. Each participant belongs to a department (5 in total), so the network has community structure.
	\item \textbf{Workplace (WP2)} Second deployment of WP, same details as WP, but larger number of participants (217) and more departments included (12). Each participant belongs to a department, so the network has community structure.
    \item \textbf{SFHH Conference (SFHH)} face-to-face interactions (Rfid sensors, range of 1.5-2 m, time resolution of 20s) of 403 participants to the 2009 SFHH conference in Nice, France (June 4-5, 2009). 
    \item \textbf{Primary School (PS)} face-to-face interactions (Rfid sensors, range of 1.5-2 m, time resolution of 20s) of 242 individuals (232 children and 10 teachers) in a primary school in Lyon, France during two days in October 2009. Each kid or teacher belongs to a class, so the network has community structure.
    \item \textbf{High school 2012 (HS2012)} face-to-face interactions (Rfid sensors, range of 1.5-2 m, time resolution of 20s) of 180 students of five classes of a high school in Marseilles, France, during 7 days (from a Monday to the Tuesday of the following week) in Nov. 2012. Each student belongs to a class, so the network has community structure.
    \item \textbf{High school 2013 (HS2013)} face-to-face interactions (Rfid sensors, range of 1.5-2 m, time resolution of 20s) of 327 students of nine classes of a high school in Marseilles, France, during 5 days in Dec. 2013. Each student belongs to a class, so the network has community structure.
    \item \textbf{Hospital} face-to-face interactions (Rfid sensors, range of 1.5-2 m, time resolution of 20s) between patients, patients and health-care workers (HCWs) and among HCWs in a hospital ward in Lyon, France, from Monday, December 6, 2010 at 1:00 pm to Friday, December 10, 2010 at 2:00 pm. The study included 46 HCWs and 29 patients.

\end{itemize}

\subsection{$E[\eta(\ell,\ell')|\mathcal{T}(\ell,\ell')<\Delta t]$ in $\mathcal{G}^2$}
\label{sec:app_dist}
\begin{figure}[H]
	\centering
	\includegraphics[width = 0.98\textwidth]{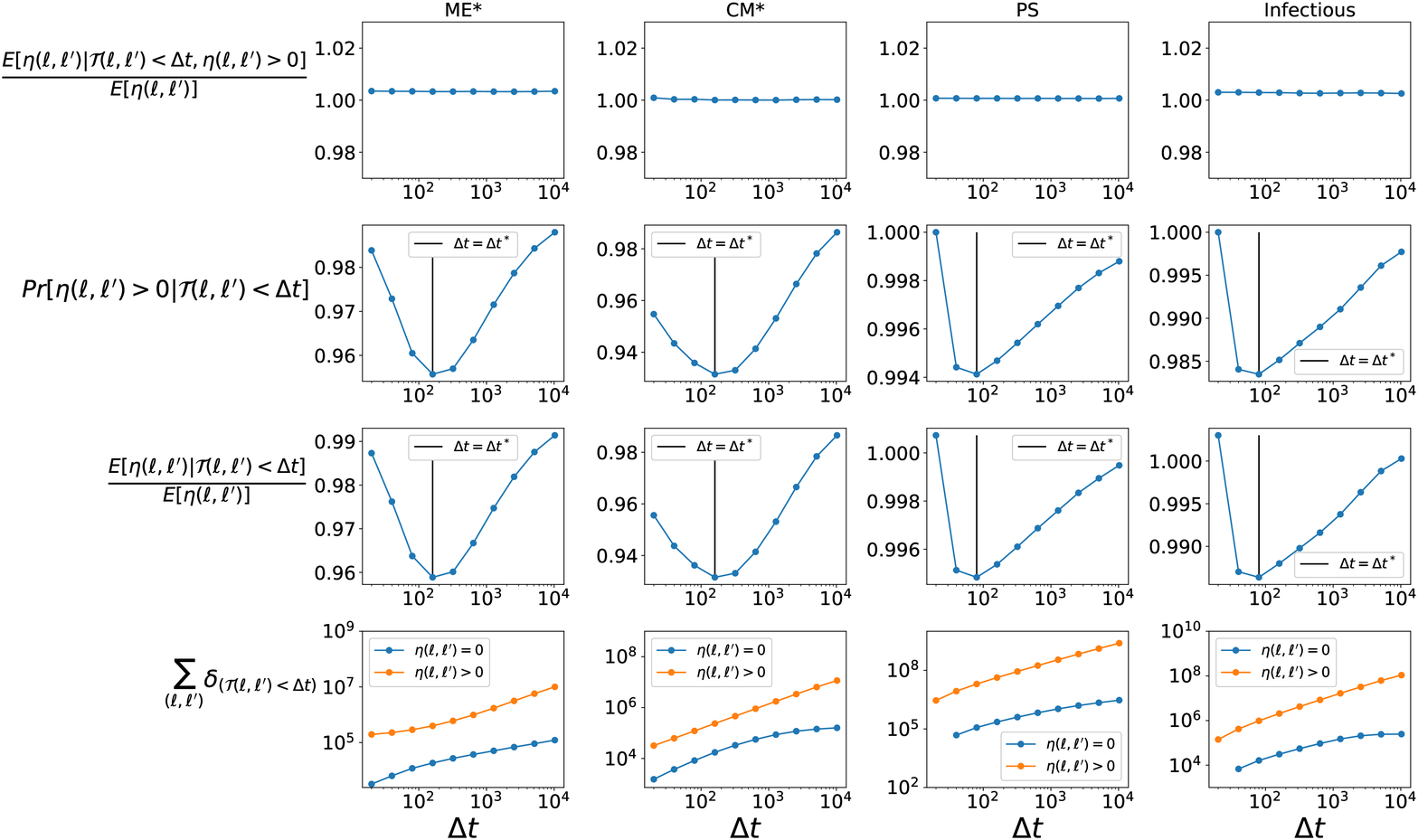}
	\caption  {The average topological distance $E[\eta(\ell,\ell') | \mathcal{T} (\ell,\ell') < \Delta t,\eta(\ell,\ell')>0 ]$ of two random contacs $\ell,\ell'$  which are not activations of the same link and have temporal distance $\mathcal{T}(\ell,\ell')<\Delta t$ (first row), the probability $Pr[\eta(\ell,\ell')>0|\mathcal{T}(\ell,\ell')< \Delta t]$ of observing two random contacts $\ell,\ell'$ which are not activations of the same link and have temporal distance $\mathcal{T}(\ell,\ell')<\Delta t$ (second row), the average topological distance  $E[\eta(\ell,\ell') | \mathcal{T} (\ell,\ell') < \Delta t ]$ of a couple of random contacts $\ell,\ell'$ with temporal distance $\mathcal{T}(\ell,\ell')<\Delta t$, and the number of couples of contacts $(\ell,\ell')$ with temporal distance $\mathcal{T}(\ell,\ell')<\Delta t$ (fourth row) and topological distance $\eta(\ell,\ell')=0$ (blue) or $\eta(\ell,\ell')>0$ (yellow)  as a function of $\Delta t$ in randomized reference model $\mathcal{G}^2$ for two examples of virtual (ME, CM) and physical (Infectious, PS) contact datasets. First and third row vertical axes are presented normalized by the average topological distance of contacts $E[\eta(\ell,\ell')]$. In second and third row the value $\Delta t = \Delta t^*$ where  $Pr[\eta(\ell,\ell')>0|\mathcal{T}(\ell,\ell')< \Delta t]$ and $E[\eta(\ell,\ell') | \mathcal{T} (\ell,\ell') < \Delta t ]$ reach their minimum is highlighted. The results are the average of 10 independent realizations of randomized network $\mathcal{G}^2$. Horizontal axes are presented in logarithmic scale. }
	\label{fig:fig12}
\end{figure}

In this subsection, we will explain the initial decreasing trend of $E[\eta(\ell,\ell')|\mathcal{T}(\ell,\ell')<\Delta t]$ with $\Delta t$ observed  in $\mathcal{G}^2$ in every considered dataset. In general, 
\begin{equation}
	E[\eta(\ell,\ell') | \mathcal{T} (\ell,\ell') < \Delta t ] = 	E[\eta(\ell,\ell') | \mathcal{T} (\ell,\ell') < \Delta t,\eta(\ell,\ell')>0 ]Pr[\eta(\ell,\ell')>0|\mathcal{T}(\ell,\ell')<\Delta t]
	\label{eq:app_eq}
\end{equation} 
where  $E[\eta(\ell,\ell') | \mathcal{T} (\ell,\ell') < \Delta t, \eta(\ell,\ell')>0]$] is the  average topological distance of couple of contacts $\ell,\ell'$ that are not activations of the same link ($\eta(\ell,\ell')>0$), given that their temporal distance $\mathcal{T}(\ell,\ell')<\Delta t$.
In $\mathcal{G}^2$, $E[\eta(\ell,\ell') | \mathcal{T} (\ell,\ell') < \Delta t,\eta(\ell,\ell')>0 ] \approx E[\eta(\ell,\ell')]$, i.e., the average topological distance of a couple of random contacts $\ell,\ell'$ which are not activations of the same link and have temporal distance $\mathcal{T}(\ell,\ell')<\Delta t$ does not depend on $\Delta t$, as shown in Figure \ref{fig:fig12}. By substituting this approximation in Equation \ref{eq:app_eq}, we obtain $E[\eta(\ell,\ell') | \mathcal{T} (\ell,\ell') < \Delta t]\approx E[\eta(\ell,\ell')] Pr[\eta(\ell,\ell')>0|\mathcal{T}(\ell,\ell')< \Delta t]$. As shown in Figure \ref{fig:fig12}, $Pr[\eta(\ell,\ell')>0|\mathcal{T}(\ell,\ell')< \Delta t]$  and $E[\eta(\ell,\ell') | \mathcal{T} (\ell,\ell') < \Delta t]$ as a function of $\Delta t$ follow the same trend and obtain the minimum at the same value of $\Delta t$. This is likely due to the relatively bursty activation patterns of single links, i.e., the high chance of observing small inter-event times in the time series of activations of single links in the considered temporal networks (see Figures \ref{fig:fig15} and \ref{fig:fig16}))

\subsection{Global probability distribution of inter-event times}
\begin{figure}[H]
\centering
\includegraphics[width = 0.9\textwidth]{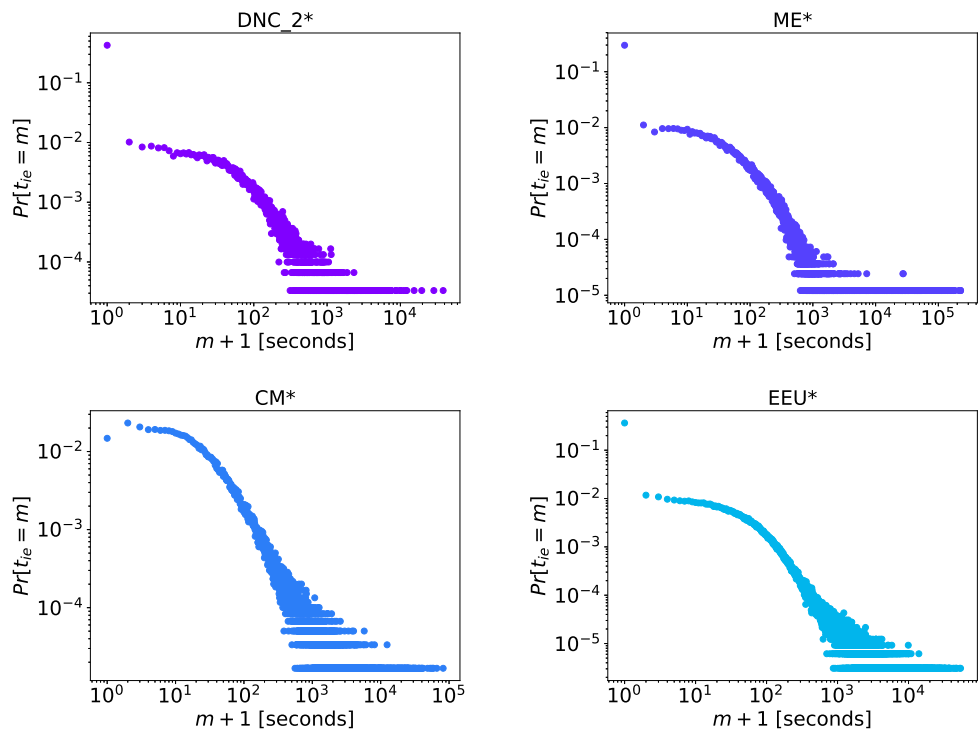}
\caption  {Probability distribution $Pr[t_{ie} = m]$ of the inter-event time of the global activity of virtual contact temporal networks. Inter-event times are reported in seconds.}
\label{fig:fig13}
\end{figure}

\begin{figure}[H]
\centering
\includegraphics[width = 0.9\textwidth]{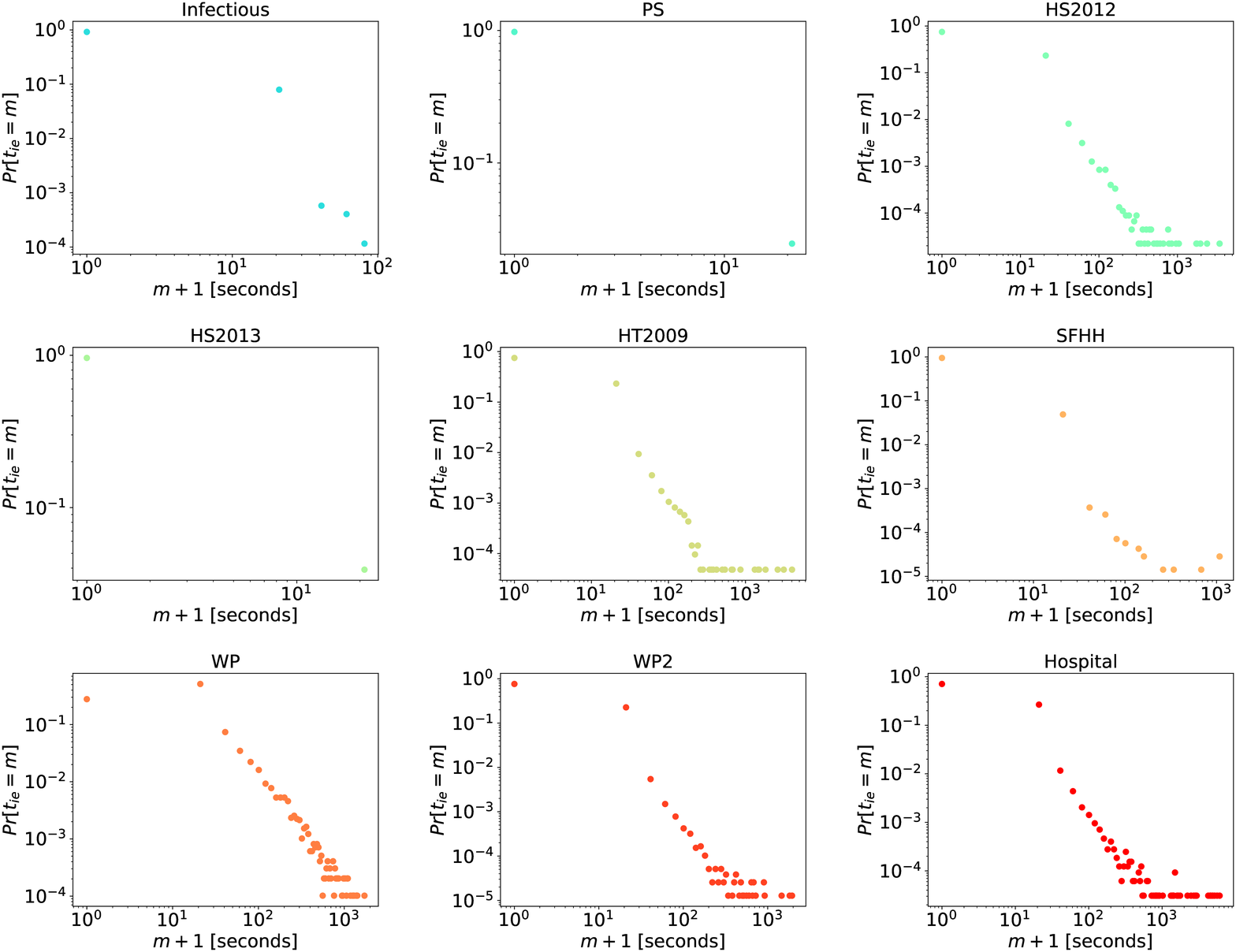}
\caption  {Probability distribution $Pr[t_{ie} = m]$ of the inter-event time of the global activity of physical contact temporal networks. Inter-event times are reported in seconds.}
\label{fig:fig14}
\end{figure}

\subsection{Inter event time distribution of links}

\begin{figure}[H]
    \centering
    \includegraphics[width = 0.9\textwidth]{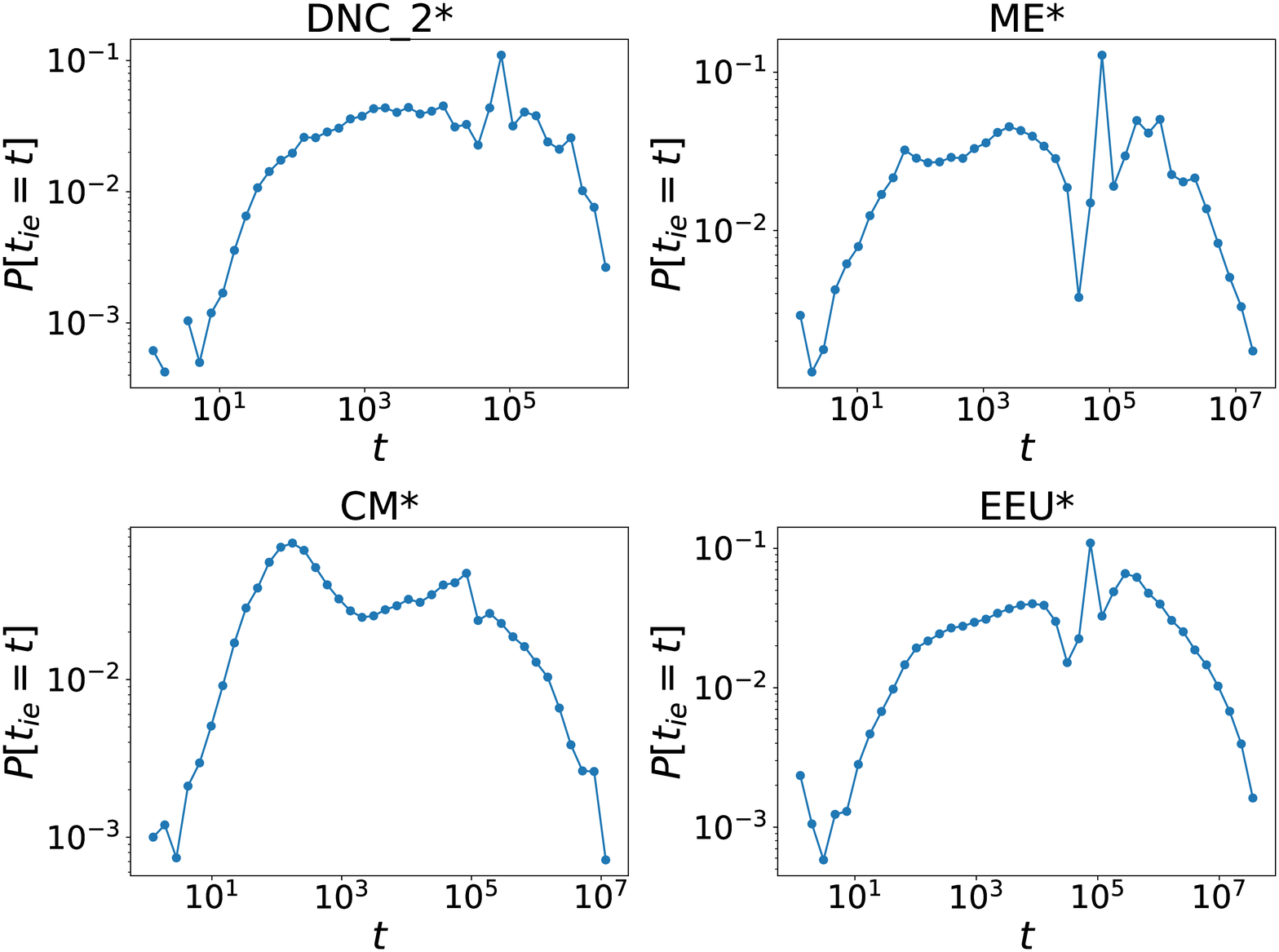}
    \caption{Inter-event time distribution of single link activity of virtual contact datasets. Note that the horizontal and vertical axes are presented in logarithmic scales. Inter-event times are measured in seconds. In total 40 logarithmic bins are split within the interval $[t_{min},t_{max}]$ where $t_{min}$ and $t_{max}$ are, respectively, the minumum and maximum inter-event time observed in the considered dataset.}
    \label{fig:fig15}
\end{figure}

\begin{figure}[H]
    \centering
    \includegraphics[width = 0.9\textwidth]{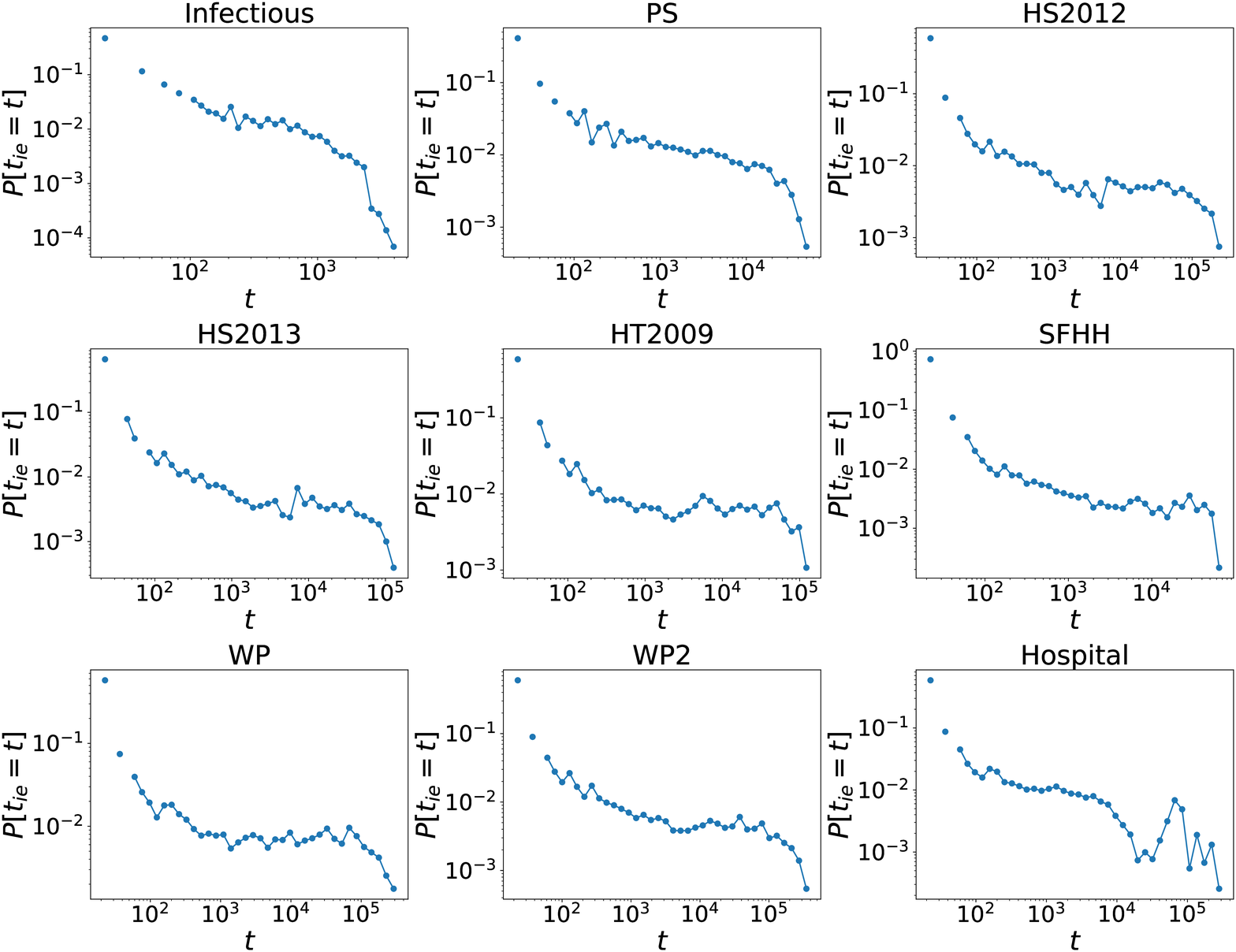}
    \caption{Inter-event time distribution of single link activity of physical contact datasets. Note that the horizontal and vertical axes are presented in logarithmic scales. Inter-event times are measured in seconds. In total 40 logarithmic bins are split within the interval $[t_{min},t_{max}]$ where $t_{min}$ and $t_{max}$ are, respectively, the minumum and maximum inter-event time observed in the considered dataset.}
    \label{fig:fig16}
\end{figure}

\subsection{Temporal correlation of local events, additional figures}

\begin{figure}[H]
    \centering
    \includegraphics[width = 0.9\textwidth]{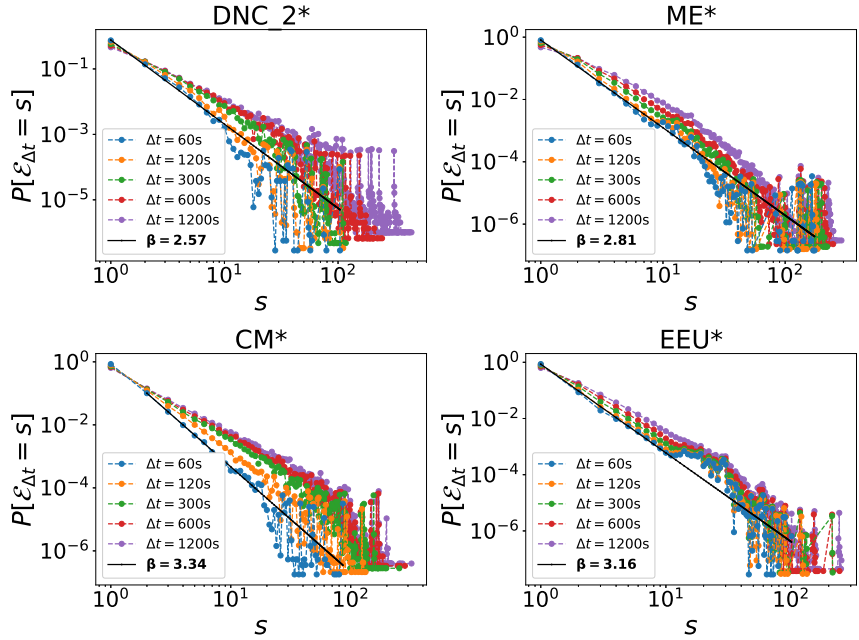}
    \caption{Train size distributions of ego network activity of $\mathcal{G}$ of virtual contact datasets with $\Delta t = $ 60 (blue),  120 (red), 300 (green), 600 (yellow), 1200 (purple) seconds. The black solid line represents the fit $P[\mathcal{E}_{\Delta t} = s] \sim s^{-\beta}$  to the distribution of the train size of $\mathcal{G}$ with $\Delta t =60s$. The power law fit and its fitting region were computed with Clauset's method  \citep{clauset2009power}. If the goodness of the power-law fit is significantly better than the exponential fit (likelihood ratio test with p-value $p<0.05$), the value of $\beta$ is reported in bold characters.}
    \label{fig:fig17}
\end{figure}

\begin{figure}[H]
    \centering
    \includegraphics[width = 0.9\textwidth]{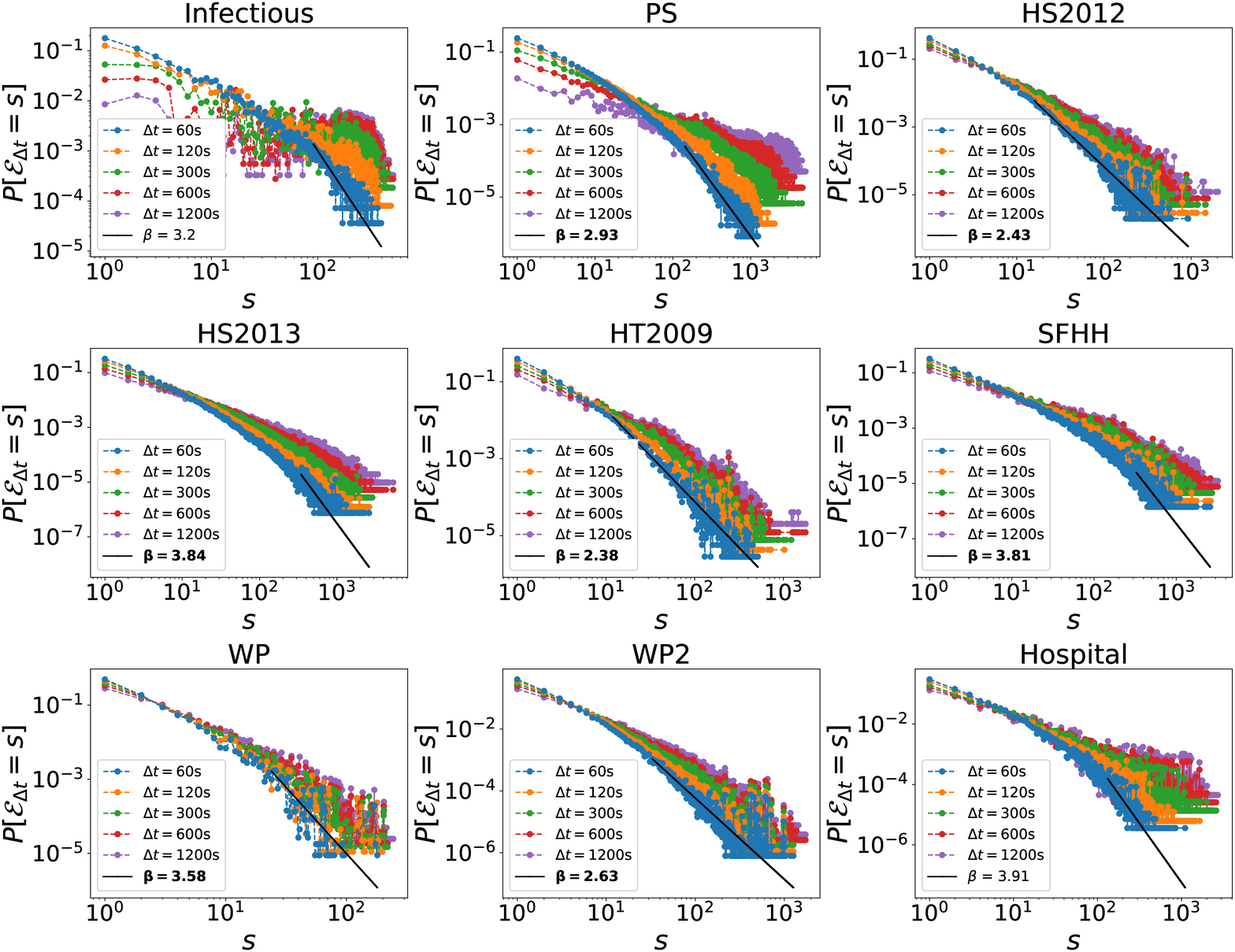}
    \caption{Train size distributions of ego network activity of $\mathcal{G}$ of physical contact datasets with $\Delta t = $ 60 (blue),  120 (red), 300 (green), 600 (yellow), 1200 (purple) seconds. The black solid line represents the fit $P[\mathcal{E}_{\Delta t} = s] \sim s^{-\beta}$  to the distribution of the train size of $\mathcal{G}$ with $\Delta t =60s$. The power law fit and its fitting region were computed with Clauset's method  \citep{clauset2009power}. If the goodness of the power-law fit is significantly better than the exponential fit (likelihood ratio test with p-value $p<0.05$), the value of $\beta$ is reported in bold characters.}
    \label{fig:fig18}
\end{figure}

\begin{figure}[H]
    \centering
    \includegraphics[width = 0.9\textwidth]{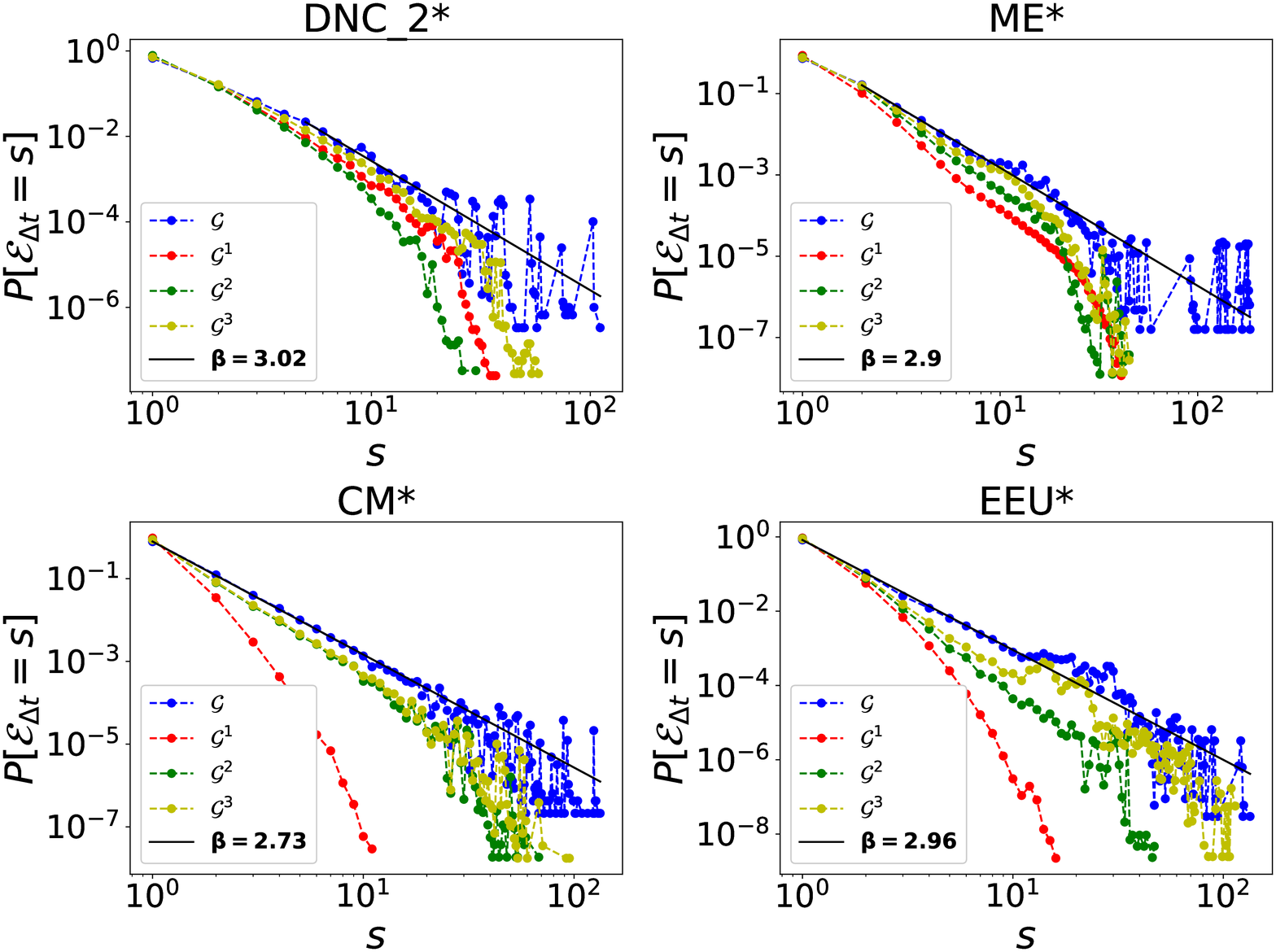}
    \caption{Train size distribution ($\Delta t = 120s$) of ego network activity for $\mathcal{G}$ (blue),  $\mathcal{G}_1$ (red),  $\mathcal{G}_2$ (green),  $\mathcal{G}_3$ (yellow) of virtual contact datasets.
    The black solid line represents the fit $P[\mathcal{E}_{\Delta t} = s] \sim s^{-\beta}$  to the distribution of the train size of $\mathcal{G}$  with $\Delta t=120s$. The power law fit and its fitting region were computed with Clauset's method  \citep{clauset2009power}. If the goodness of the power-law fit is significantly better than the exponential fit (likelihood ratio test with p-value $p<0.05$), the value of $\beta$ is reported in bold characters.}
    \label{fig:fig19}
\end{figure}

\begin{figure}[H]
    \centering
    \includegraphics[width = 0.9\textwidth]{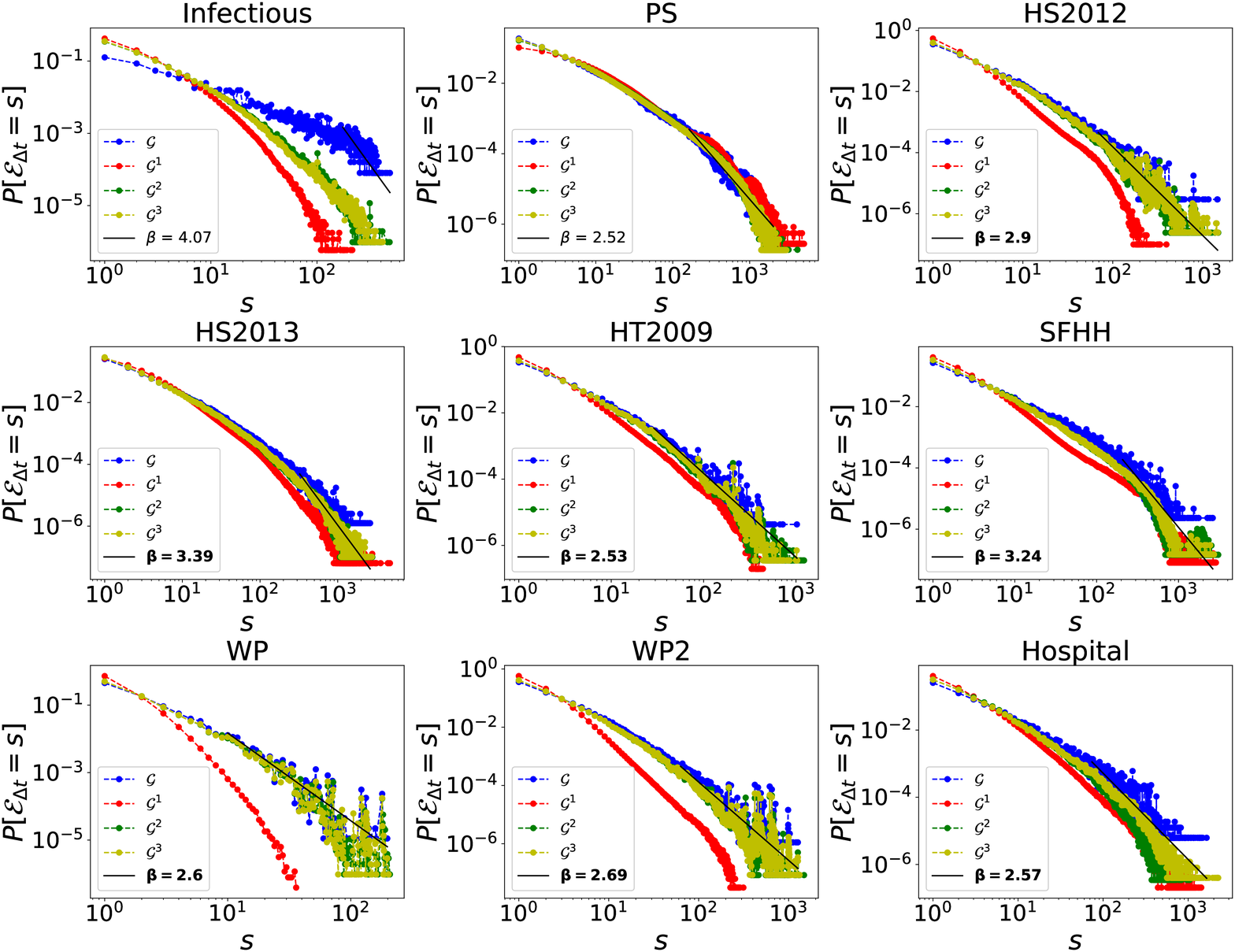}
    \caption{Train size distribution ($\Delta t = 120s$) of ego network activity for $\mathcal{G}$ (blue),  $\mathcal{G}_1$ (red),  $\mathcal{G}_2$ (green),  $\mathcal{G}_3$ (yellow) of physical contact datasets. The black solid line represents the fit $P[\mathcal{E}_{\Delta t} = s] \sim s^{-\beta}$  to the distribution of the train size of $\mathcal{G}$  with $\Delta t = 120$. The power law fit and its fitting region were computed with Clauset's method  \citep{clauset2009power}. If the goodness of the power-law fit is significantly better than the exponential fit (likelihood ratio test with p-value $p<0.05$), the value of $\beta$ is reported in bold characters.}
    \label{fig:fig20}
\end{figure}

\begin{figure}[H]
\centering
\includegraphics[width = 0.9\textwidth]{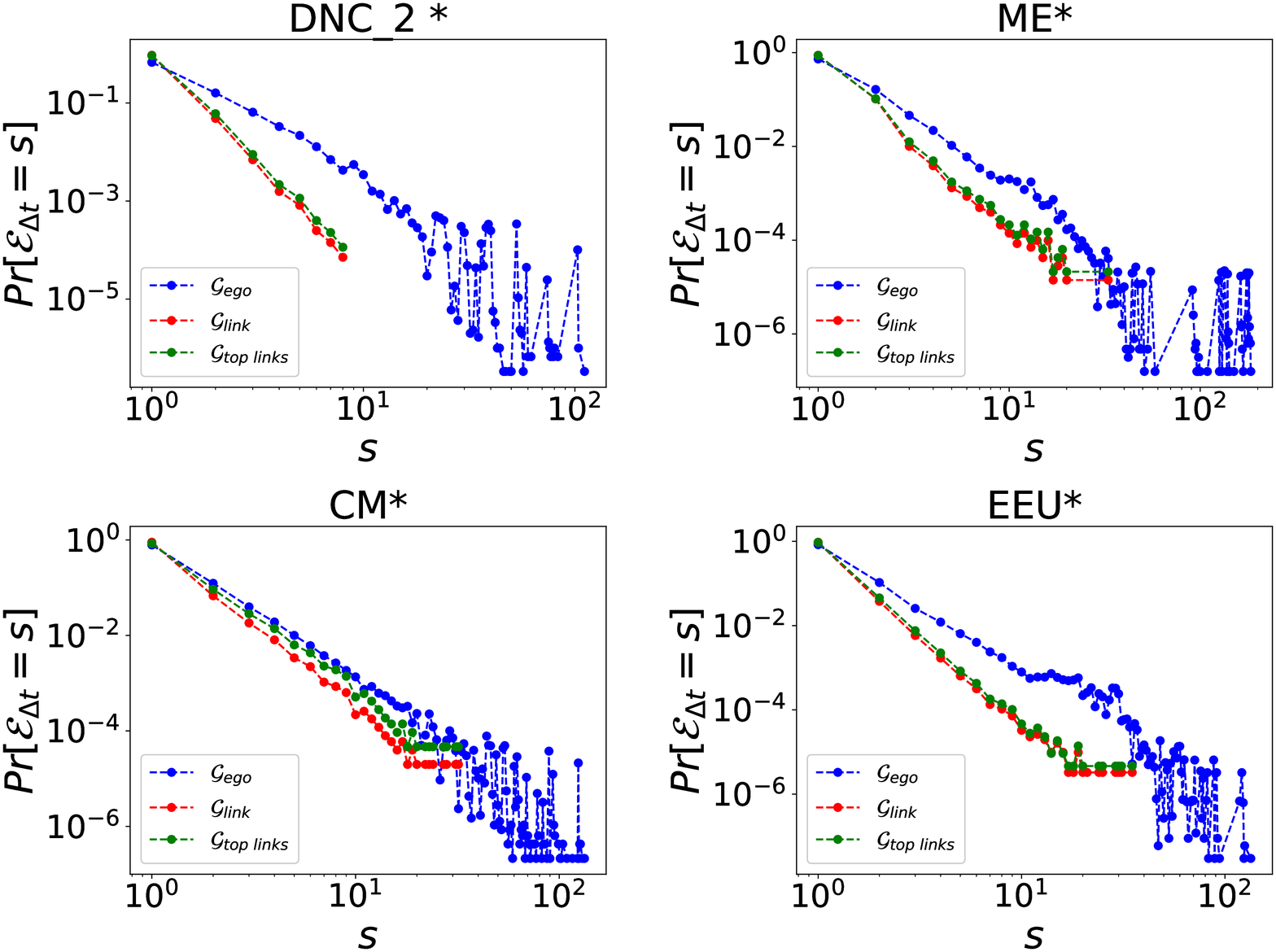}
\caption{Train size distribution ($\Delta t = 120s$) of ego network activity (blue),  single link activity (red),  most active link activity (green)  of virtual contact datasets. Note that the horizontal and vertical axes are presented in logarithmic scales.}
\label{fig:fig21}
\end{figure}

\begin{figure}[H]
\centering
\includegraphics[width = 0.9\textwidth]{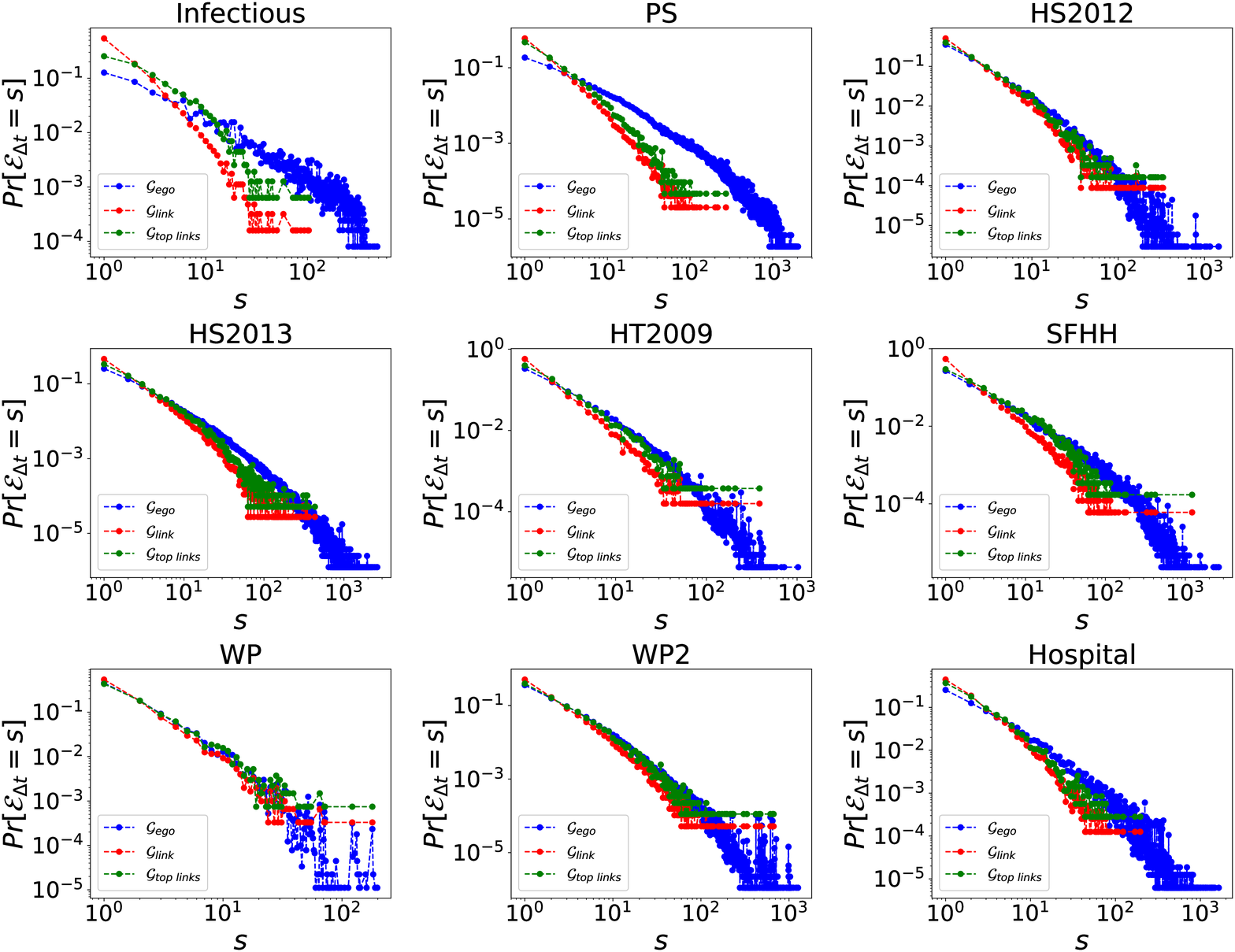}
\caption{Train size distribution ($\Delta t = 120s$) of ego network activity (blue),  single link activity (red),  most active link activity (green)  of physical contact datasets. Note that the horizontal and vertical axes are presented in logarithmic scales.}
\label{fig:fig22}
\end{figure}

\begin{figure}[H]
\centering
\includegraphics[width = 0.9\textwidth]{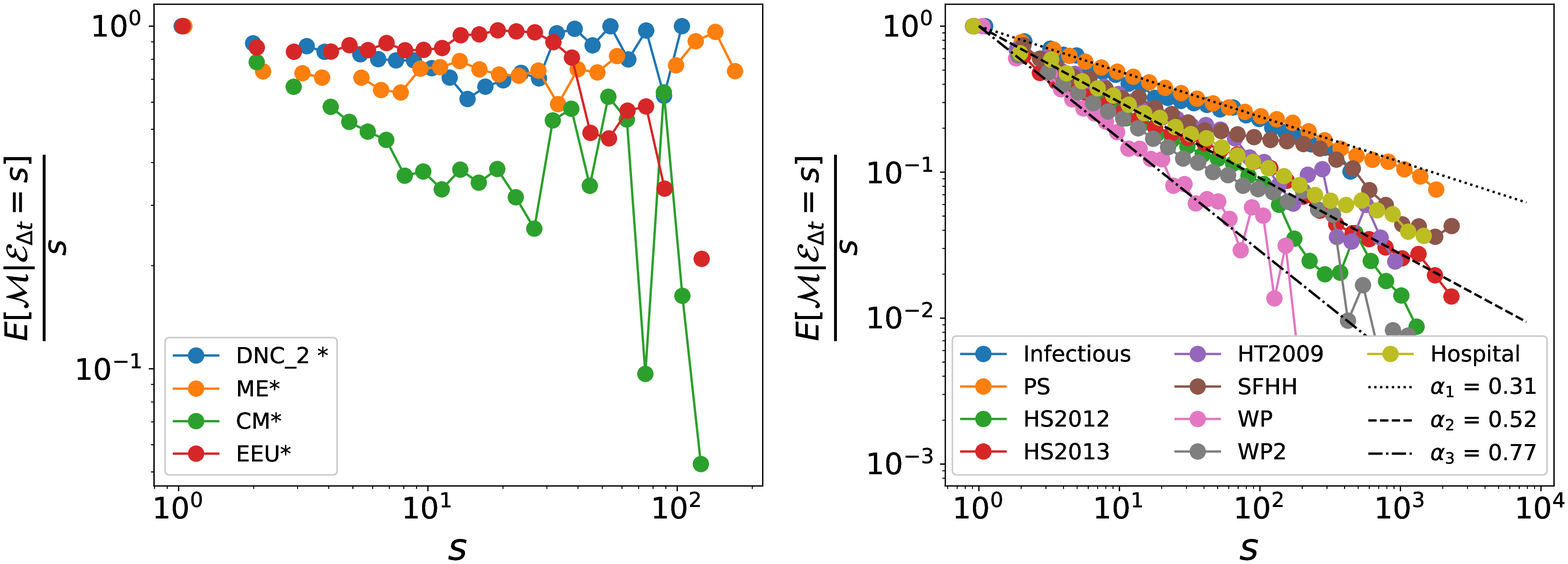}
\caption  {The average number of active links $\frac{E[\mathcal{M}|\mathcal{E}_{\Delta t}=s]}{s}$ for trains with size $\mathcal{E}_{\Delta t}=s$ ($\Delta t = 120 s$), normalized by the train size $s$ of the ego networks for virtual (left) and physical (right) contact datasets. The three reference lines in right plot indicate $\frac{E[\mathcal{M}|\mathcal{E}_{\Delta t}=s]}{s} = s^{-\alpha}$ with slope $\alpha_1 = 0.31$ (dotted), $\alpha_2 = 0.52$ (dashed) and $\alpha_3 = 0.77$ (dash-dot).  Note that the horizontal and vertical axes are presented in logarithmic scales. In total 30 logarithmic bins are split within the interval $[1, s_{max}]$, where $s_{max}$ is the largest train size observed in the considered real temporal network.}
\label{fig:fig23}
\end{figure}

\end{document}